\documentclass[number,preprint,3p]{elsarticle}

\usepackage{multirow}
\usepackage{color}
\usepackage{graphicx}
\usepackage{subcaption}
\usepackage{algorithm}
\usepackage{bm}
\usepackage[colorlinks]{hyperref}
\usepackage{amssymb}
\usepackage{amsthm}
\usepackage{amsmath}
\usepackage{booktabs}
\usepackage{url}
\usepackage{scrextend}
\usepackage{epstopdf}
\usepackage{float}
\usepackage{tablefootnote}
\usepackage{mathtools}
\usepackage{soul}
\usepackage[perpage]{footmisc}
\usepackage{lineno}
\makeatletter
\newenvironment{breakablealgorithm}
  {
   \begin{center}
     \refstepcounter{algorithm}
     \hrule height.8pt depth0pt \kern2pt
     \renewcommand{\caption}[2][\relax]{
       {\raggedright\textbf{\ALG@name~\thealgorithm} ##2\par}%
       \ifx\relax##1\relax 
         \addcontentsline{loa}{algorithm}{\protect\numberline{\thealgorithm}##2}%
       \else 
         \addcontentsline{loa}{algorithm}{\protect\numberline{\thealgorithm}##1}%
       \fi
       \kern2pt\hrule\kern2pt
     }
  }{
     \kern2pt\hrule\relax
   \end{center}
  }
\makeatother
\makeatletter
\gdef\urlauthor#1#2{\g@addto@macro\@elsuads{\let\corref\@gobble%
     \def\@@tmp{#1}\raggedright\eadsep
     {\ttfamily\url{\expandafter\strip@prefix\meaning\@@tmp}}\space(#2)%
     \def\eadsep{\unskip,\space}}%
}
\gdef\emailauthor#1#2{\stepcounter{ead}%
     \g@addto@macro\@elseads{\raggedright%
      \let\corref\@gobble\def\@@tmp{#1}%
      \eadsep{\ttfamily\href{mailto:\expandafter\strip@prefix\meaning\@@tmp}{\expandafter\strip@prefix\meaning\@@tmp}}
      (#2)\def\eadsep{\unskip,\space}}%
}
\makeatother

\def\r{\mathbb{R}}
\def\rn{\mathbb{R}^n}

\newcommand{\vect}[1]{\boldsymbol{#1}}

\makeatletter
\let\@afterindenttrue\@afterindentfalse
\makeatother

\biboptions{numbers,sort&compress,square}
\journal{Reliability Engineering \& System Safety}

\makeatletter
\patchcmd{\ps@pprintTitle}
  {Preprint submitted to}
  {Accepted manuscript in}
  {}{}
\makeatother

\begin{document}
\begin{frontmatter}
\renewcommand{\thefootnote}{\fnsymbol{footnote}}



\title{Dimensionality reduction can be used as a surrogate model for high-dimensional forward uncertainty quantification}

 \author[1]{Jungho Kim}
 \author[2]{Sang-ri Yi}
 \author[3]{Ziqi Wang\corref{cor1}}
\ead{ziqiwang@berkeley.edu}  
         \cortext[cor1]{Corresponding author}
 \address[1]{Department of Civil and Environmental Engineering, Sejong University, South Korea}

  \address[2]{Department of Civil and Environmental Engineering, Rice University, United States}
         
 \address[3]{Department of Civil and Environmental Engineering, University of California, Berkeley, United States}
\begin{abstract}
We introduce a method to construct a stochastic surrogate model from the results of dimensionality reduction in forward uncertainty quantification. The hypothesis is that the high-dimensional input augmented by the output of a computational model admits a low-dimensional representation. This assumption can be met by numerous uncertainty quantification applications with physics-based computational models. The proposed approach differs from a sequential application of dimensionality reduction followed by surrogate modeling, as we ``extract" a surrogate model from the results of dimensionality reduction in the input-output space. This feature becomes desirable when the input space is genuinely high-dimensional. The predictive distribution is obtained by generating samples from a transition kernel that encodes the dimensionality reduction and a feature-space conditional distribution. The resulting surrogate model operates as a stochastic simulator that propagates a deterministic input into a stochastic output, preserving the convenience of a sequential ``dimensionality reduction + Gaussian process regression" approach while overcoming some of its limitations. The proposed method is demonstrated through three uncertainty quantification problems characterized by high-dimensional input uncertainties. 
\end{abstract}

\end{frontmatter}
\renewcommand{\thefootnote}{\fnsymbol{footnote}}

\section{Introduction}

\noindent Forward uncertainty quantification (UQ) aims to quantify the output uncertainty propagated from the uncertain input of a computational model. This subject finds increasingly broad applications in computational science and engineering and is evolving rapidly alongside the unprecedented growth of data science and machine learning techniques. Efficient forward UQ for complex computational models remains an outstanding challenge, even with the expansion of computational resources \cite{marelli2021stochastic,qin2023machine}. This challenge becomes more pronounced when the models require high-fidelity simulations using black-box solvers. Moreover, the quantities of interest in forward UQ analysis, such as probabilities of critical events, quantiles, and statistical moments, are often determined by integrals over a high-dimensional space of input parameters, which suffer from the curse of dimensionality \cite{wang2016cross,tripathy2018deep,kim2021quantile,kontolati2022survey,zhou2023partial}.

Monte Carlo simulation (MCS) can be easily applied to black-box models and is insensitive to the input dimensionality, but it suffers from a slow convergence rate. Surrogate modeling becomes a key strategy in this context: if an expensive computational model can be effectively replaced by an efficient surrogate, UQ analysis and other outer-loop applications can be significantly accelerated. In recent studies of UQ methods \cite{schobi2017uncertainty,kim2020probabilistic,kim2020probability,zheng2024learnable}, the polynomial chaos expansion, Gaussian process/Kriging, and neural network are among the most popular surrogate models due to their flexibility in fitting generic nonlinear functions. However,these surrogate models face difficulties in dealing with high-dimensional uncertainties. The volume of a Euclidean space expands exponentially with its dimensionality; accordingly, the parametric space and the amount of training data for surrogate modeling also increase significantly. This property appears to impose a fundamental constraint on the scalability of surrogate modeling. However, this work and many real-world UQ applications are not focused on some ``generic" high-dimensional problems. Instead, the computational models are often governed and constrained by physical laws, which may admit low-dimensional representations. 

In the context of UQ problems with high-dimensional input uncertainties, it is promising to exploit various linear and nonlinear dimensionality reduction, manifold learning, and model order reduction techniques. A straightforward, sequential approach to combining dimensionality reduction with surrogate modeling involves: (1) identifying a low-dimensional feature space representation for the high-dimensional input, using information of the input only, and (2) building a surrogate computational model in this feature space. Recent studies have explored the use of principal component analysis \cite{lataniotis2020extending,liu2021global}, diffusion map \cite{kalogeris2020diffusion}, Grassmann manifold \cite{giovanis2018uncertainty,giovanis2020data}, and multipoint nonlinear kernel-based manifold \cite{dos2022grassmannian} in constructing the feature mapping for high-dimensional input or high-dimensional output; clustering \cite{kontolati2022manifold} and interpolation \cite{giovanis2020data,kubicek2015high} techniques have been applied to improve the prediction accuracy. While a sequential combination of dimensionality reduction and surrogate modeling can be effective for some applications, it faces considerable limitations in the presence of genuinely high-dimensional input uncertainties. A notable example is the stochastic dynamics analysis of structural systems subjected to wide-band excitation, which is frequently encountered in earthquake engineering \cite{kim2024adaptive,wang2024probabilistic}. Recent studies have proposed advanced strategies combining dimensionality reduction and surrogate modeling, such as convolutional neural networks for spatial uncertainty \cite{shi2024convolutional}, autoencoder-based time-dependent latent learning \cite{zhang2024rLSTM}, sensitivity-guided \cite{chen2024dimensionality} or physics-informed \cite{zhou2022uncertainty} metamodeling for UQ.

An alternative to the sequential combination is a supervised approach that integrates dimensionality reduction with feature space surrogate modeling. Specifically, a supervised approach can take the form of (a) optimizing the dimensionality reduction and surrogate modeling jointly using samples of the input-output pairs \cite{lataniotis2020extending}, possibly guided by active learning \cite{kim2024adaptive}, (b) introducing dimensionality reduction as an inherent construction of surrogate models, either through sparsity constraints for data-fitting models \cite{konakli2016polynomial,giovanis2020data} or through projection bases for reduced order models \cite{schilders2008model,chinesta2011short}, or (c) merging dimensionality reduction with surrogate modeling by uncovering low-dimensional probabilistic models in the product space of input and output parameters \cite{soize2016data,soize2019entropy,soize2020physics}.  

The approach (c), termed Probabilistic Learning on Manifolds (PLoM), is appealing because it does not require a differentiation between dimensionality reduction and surrogate modeling. This distinction can be redundant for ``high-dimensional input and low-dimensional output" problems, because one can argue the ``optimal" dimensionality reduction is the computational model itself, coinciding with the ``optimal" surrogate model. PLoM performs predictions by generating samples concentrated on the embedded manifold of the input-output space, while also satisfying constraints on input variables \cite{soize2020physics,zhong2023surrogate}. This makes the usage of PLoM different from the traditional surrogate models that predict the output given a precise input value. During the constrained resampling process, PLoM requires a reconstruction mapping from the low-dimensional representation to the original input-output space. This requirement limits the applicability of general nonlinear dimensionality reduction techniques, particularly when a reliable reconstruction is unavailable or computationally expensive. Recent advances such as autoencoders and other machine-learning-based reconstruction techniques \cite{shi2024convolutional,zhang2024rLSTM} partially address this issue; however, the proposed method circumvents the reconstruction step entirely, thereby enabling broader applicability to nonlinear dimensionality reduction techniques.

This paper develops a method to ``extract" a stochastic surrogate model from the results of dimensionality reduction. The method proceeds as follows: (i) perform dimensionality reduction in the input-output space of a computational model; (ii) construct a conditional distribution that connects the low-dimensional features with the model output; (iii) create a stochastic simulator for predicting the model output from the high-dimensional input. This simulator is extracted via a transition kernel that encodes a fixed-point iteration process, requiring no input-specific training. The proposed method avoids the need for a reconstruction mapping, thereby ensuring compatibility with a broad range of nonlinear dimensionality reduction techniques while maintaining computational simplicity. In addition, the method benefits from input–output physics-based learning in the reduced space, enabling accurate treatment of complex, high-dimensional uncertainty structures, including irregular or highly non-smooth features. Furthermore, the core of the method, step (iii), does not introduce any additional parameters to tune, since the hyperparameters of the dimensionality reduction and conditional distribution model are determined during the training phase and remain fixed during the prediction stage. This allows the method to serve as an easy-to-implement baseline for studying more sophisticated high-dimensional surrogate modeling techniques. As a result, the approach provides a stochastic simulator that propagates deterministic inputs into probabilistic outputs, reflecting the inherent uncertainty in complex models.

Section \ref{Background} states the problem and provides a brief review of the sequential ``dimensionality reduction + surrogate modeling" approach and PLoM. Section \ref{Overview} introduces the general formulations of the proposed method. Section \ref{DRSMdetails} presents the computational details. Section \ref{Examples} demonstrates the performance of the proposed method with numerical examples. Section \ref{Remarks} discusses limitations and future research directions. The paper concludes with a summary in Section \ref{Conclusion}.

\section{Problem statement} \label{Background}
Consider a system with input random variables $\vect{X}\in\rn$ representing the source of randomness, and the output $Y\in\r$ characterizing the performance of interest. The computational model is represented by $\mathcal{M}: \vect{x}\mapsto y$. Given the joint probability distribution of $\vect{X}$, we seek the probability distribution and generalized moments of $Y$. This task is challenging because the model $\mathcal{M}$ may involve computationally intensive simulations, and the input $\vect{X}$ can be high-dimensional.

\subsection{Sequential combination of dimensionality reduction with surrogate modeling}
A straightforward, sequential combination of dimensionality reduction with surrogate modeling first identifies a low-dimensional representation of the input $\vect{X}$, and then constructs a surrogate model in the low-dimensional feature space. Specifically, the dimensionality reduction step constructs the forward and inverse projections:
\begin{equation}  \label{Eq:DR}
\begin{aligned}
&\mathcal{H}\left(\vect{x};\vect{\theta}_{\mathcal{H}}\right):\vect x\in\rn\mapsto \vect{\vect\psi}\in\r^{d}\,,\\
&\mathcal{H}^{-1}\left(\vect{\vect\psi};\vect{\theta}_{\mathcal{H}}\right):\vect{\vect\psi}\in\r^{d}\mapsto\hat{\vect{x}}\in\rn\,,  
\end{aligned}
\end{equation}
where $\vect{\vect\psi}$ is the low-dimensional feature vector, $\hat{\vect{x}}$ is the reconstructed input vector from the inverse mapping $\mathcal{H}^{-1}$, and $\vect{\theta}_{\mathcal{H}}$ denotes parameters that characterize the dimensionality reduction. Examples of dimensionality reduction methods include principal component analysis (PCA), kernel-PCA, diffusion maps, Isomap, locally linear embedding, and autoencoder (see \cite{van2009dimensionality,kontolati2022survey} for a review). Note that the inverse projection is often approximate and may involve additional assumptions \cite{kalogeris2020diffusion,kontolati2022manifold}. Depending on the specific design of the ``dimensionality reduction + surrogate modeling" approach, the inverse mapping $\mathcal{H}^{-1}$ may not be necessary. For example, if a pre-specified pool of samples are used in predicting the quantities of interest, the $\vect x$ and $\vect\psi$ have point-wise bijective mapping, and an analytic inversion $\mathcal{H}^{-1}$ can be circumvented (see \cite{kim2024adaptive} for a concrete example).

Provided with the dimensionality reduction, the surrogate model $\hat{\mathcal{M}}$ maps the feature vector $\vect\psi$ into an approximation of $y$, i.e., 
\begin{equation} \label{Eq:seq_DR_SM}
\hat{y} = \hat{\mathcal{M}}\left(\vect{\vect\psi};\vect{\theta}_{\hat{\mathcal{M}}}\right) = \hat{\mathcal{M}}\left(\mathcal{H}\left(\vect{x};\vect{\theta}_{\mathcal{H}}\right);\vect{\theta}_{\hat{\mathcal{M}}}\right)\,,
\end{equation}
where $\vect{\theta}_{\hat{\mathcal{M}}}$ denotes parameters of the surrogate model. The surrogate $\hat{\mathcal{M}}$ is typically constructed by standardized parametric models, such as Gaussian process and polynomial chaos expansion, which are known to be effective for low-dimensional problems. Consequently, the accuracy of Eq.~\eqref{Eq:seq_DR_SM} is dominated by the feature mapping $\mathcal{H}$. If an unsupervised approach is adopted to construct $\mathcal{H}$, meaning the computational model $\mathcal{M}$ is not used in formulating $\mathcal{H}$, the ``dimensionality reduction + surrogate modeling" approach would only work for specialized problems. An example of such a specialized problem is when the components of $\vect X$ exhibit strong correlations. This limitation motivates the development of supervised and coupled approaches. 

\subsection{Probabilistic learning on manifolds (PLoM)}
The aforementioned limitation is lifted when dimensionality reduction is performed in a supervised setting, utilizing information from the model $\mathcal{M}$. One such approach is PLoM, where a lower-dimensional manifold is identified from the augmented vector of input and output, denoted by $ \vect{z}\equiv[ \vect{x}, y]$. In this context, the concept of dimensionality reduction overlaps with that of surrogate modeling, because the learning of manifold involves identifying probable locations of input-output pairs, which closely resembles the objective of surrogate modeling. The latent space projection is expressed as:
\begin{equation}  \label{Eq:PLoM1} 
\mathcal{H}\left(\vect{z};\vect{\theta}_{\mathcal{H}}\right):\vect{z}\in\r^{n+1}\mapsto\vect\psi_{\vect z}\in\r^d\,,
\end{equation}
where $\vect\psi_{\vect z}$ is an underlying low-dimensional representation of $\vect{z}$ with probability distribution $f_{\vect\Psi_{\vect z}}$. In PLoM, the prediction operates differently from the conventional surrogate models. It begins by approximating and generating samples from the conditional distribution $f_{{\vect\Psi}_{\vect{z}}|\vect X_c}(\vect{\vect\psi}_{\vect{z}}|\vect{x}\in \vect{X}_c)$, where $\vect{X}_c$ is a set of realizations of $\vect{X}$ that satisfies some desired conditions. The elements of $\vect{X}_c$ can be generated from a distribution with specified mean and variance \cite{zhong2023surrogate,soize2020physics}, or they can be predefined deterministic values \cite{soize2018probablistic}. The samples in the feature space are then transformed into the original input-output space through the inverse projection: 
\begin{equation}  \label{Eq:PLoM2}
\mathcal{H}^{-1}\left(\vect{\hat{\vect\psi}_{\vect{z}}};\vect{\theta}_{\mathcal{H}}\right):\vect{\hat{\vect\psi}_{\vect{z}}}\in\r^d\mapsto\hat{\vect{z}}\equiv(\hat{\vect x},\hat{y})\in\r^{n+1}\,,
\end{equation}
where ``hats" are used to emphasize that the feature space samples from an approximate $f_{{\vect\Psi}_{\vect{z}}|\vect X_c}$ as well as the reconstructed $\vect z$ contain errors. Provided with effective sampling and inverse projection algorithms, the distribution of the reconstructed samples $\hat{\vect x}$ should be close to that of $\vect{X}_c$. Consequently, the samples $\hat{y}$ generated in this process can be used to predict the statistics of $\mathcal{M}(\vect x)$, $\vect x\in\vect X_c$. 

The challenge of PLoM arises from sampling the conditional distribution $f_{{\vect\Psi}_{\vect{z}}|\vect X_c}$. A commonly used technique involves solving the Itô stochastic differential equation \cite{zhong2023surrogate,soize2020physics}. However, this approach is not applicable when $\vect{X}_c$ represents a single input vector $\vect{x}$. An alternative strategy is to generate numerous samples from $f_{\vect\Psi_{\vect{z}}}(\vect{\vect\psi}_{\vect z})$ and transform them into the original input-output space for post-processing, such as performing interpolation with kernel density estimation (KDE) \cite{soize2018probablistic}. This approach may not scale well with dimension. Furthermore, requiring an inverse projection poses a strong constraint when considering the integration with various dimensionality reduction techniques.

\section{Extracting a surrogate model from dimensionality reduction}  \label{Overview}
The proposed Dimensionality Reduction-based Surrogate Modeling method (DR-SM) consists of the following three steps.  
\begin{itemize}
    \item Dimensionality reduction in the input-output space: construct $\mathcal{H}:\vect z\equiv(\vect x,y)\in\mathbb{R}^{n+1}\mapsto\vect\psi_{\vect z}\in\mathbb{R}^d$. 
    \item Construct a conditional distribution $f_{\hat Y|\vect\Psi_{\vect z}}(\hat y|\vect\psi_{\vect z})$ to predict $y$ given $\vect\psi_{\vect z}$. 
    \item Extract a stochastic surrogate model $f_{\hat Y|\vect X}(\hat y|\vect x)$ using $\mathcal{H}$ and $f_{\hat Y|\vect\Psi_{\vect z}}( \hat y|\vect\psi_{\vect z})$.
\end{itemize}
The first and second steps can be completed using existing dimensionality reduction and distribution modeling techniques. The third step is challenging and requires specialized techniques. The challenge lies in the fact that the feature vector $\vect\psi_{\vect z}$ is contributed by both $\vect x$ and $y$, and the goal is to construct a ``decoupled" surrogate model $f_{\hat Y|\vect X}(\hat y|\vect x)$ that predicts $y$ given $\vect x$. The first and second steps are referred to as the \textit{training stage} of DR-SM, while the last step is the \textit{prediction stage}.

\begin{figure}[H]
  \centering
  \includegraphics[scale=0.50] {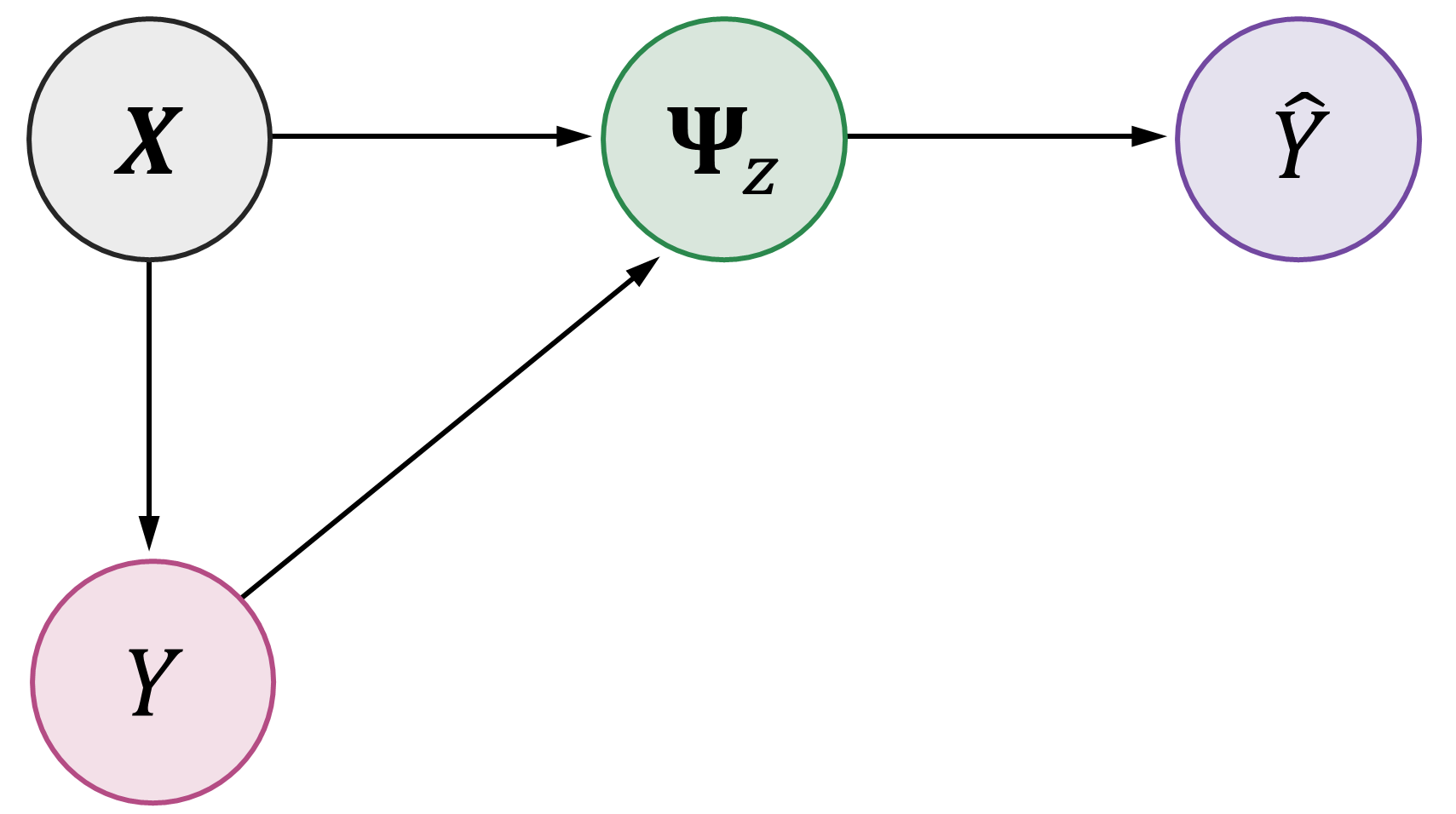}
  \caption{\textbf{The interdependencies between $\vect X$, $Y$, $\vect\Psi_{\vect z}$, and $\hat Y$ for the training stage.}}
  \label{Fig:ProbGraph1}
\end{figure}

Figure \ref{Fig:ProbGraph1} illustrates the interdependencies between $\vect X$, $Y$, $\vect\Psi_{\vect z}$, and $\hat Y$ during the training stage. The desired surrogate model $f_{\hat Y|\vect X}(\hat y|\vect x)$ can be expressed by:
\begin{equation}  \label{Eq:exactmod}
    f_{\hat Y|\vect X}(\hat y|\vect x)=\int f_{\hat Y|\vect\Psi_{\vect z}}(\hat y|\vect\psi_{\vect z})f_{\vect\Psi_{\vect z}|\vect XY}(\vect\psi_{\vect z}|\vect x y)f_{Y|\vect X}(y|\vect x)\,d\vect\psi_{\vect z}\,dy\,,
\end{equation}
where $f_{\vect\Psi_{\vect z}|\vect XY}$ is associated with the dimensionality reduction and $f_{Y|\vect X}$ with the computational model. They are Dirac delta functions if the dimensionality reduction and computational model are deterministic. In general, Eq.~\eqref{Eq:exactmod} encodes an increase in uncertainty from $f_{Y|\vect X}$ to $f_{\hat Y|\vect X}$. This is because $f_{\hat Y|\vect X}$ is susceptible to errors/uncertainties in the dimensionality reduction and the feature space conditional distribution. In the theoretically ideal case where the computational model is deterministic and both the dimensionality reduction and the $f_{\hat Y|\vect\Psi_{\vect z}}$ model are perfect, the right-hand side of Eq.~\eqref{Eq:exactmod} yields $\delta(\hat y-\mathcal{M}(\vect x))$, i.e., a perfect surrogate model. In practice, $f_{\hat Y|\vect X}$ offers an approximation to $f_{Y|\vect X}$. The accuracy hinges on the performance of the dimensionality reduction, because $f_{\hat Y|\vect\Psi_{\vect z}}$ is a low-dimensional distribution that can be effectively modeled by various statistical learning methods. Therefore, the development of this paper proceeds under the premise that the dimensionality reduction is effective, allowing us to leverage efficient compression for making predictions. This premise conceptually aligns well with the discussions on the equivalence between data compression (compressibility) and prediction (predictability) \cite{deletang2023language,vitanyi1997prediction}.

\begin{figure}[H]
  \centering
  \includegraphics[scale=0.51] {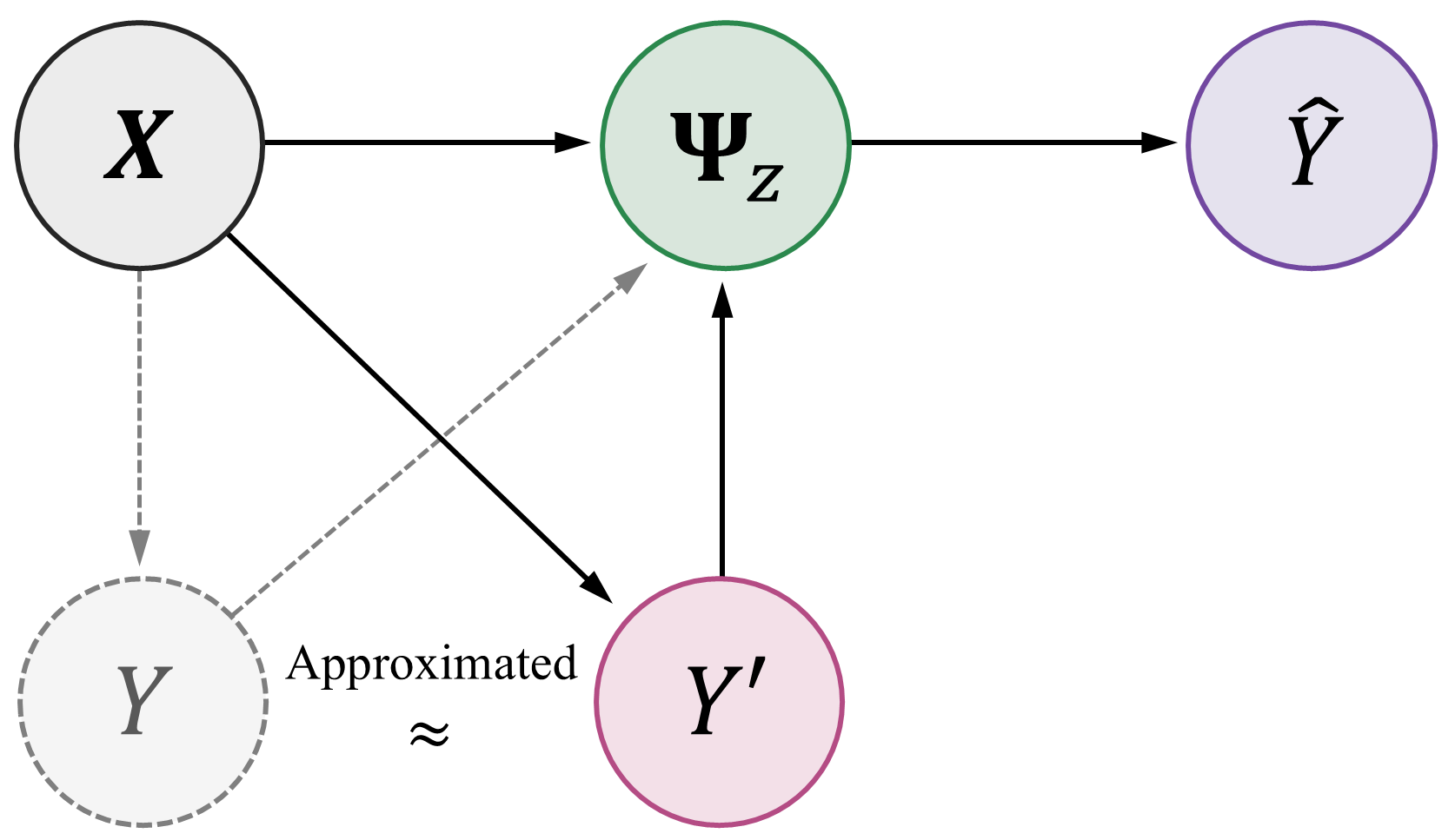}
  \caption{\textbf{The interdependencies between $\vect X$, $Y$, $\vect\Psi_{\vect z}$, and $\hat Y$ for the prediction stage.}}
  \label{Fig:ProbGraph2}
\end{figure}

Eq.~\eqref{Eq:exactmod} represents the ``true" surrogate model to be extracted from dimensionality reduction. However, this equation cannot be directly used for predictive purposes, as it depends on knowledge of the original model $f_{Y|\vect X}(y|\vect x)$ --- the goal of surrogate modeling. This creates a classic ``Catch-22" situation, necessitating a reformulation of the dependency structure shown in Figure \ref{Fig:ProbGraph1}. To address this, the conditional distribution $f_{Y|\vect X}$ in Figure \ref{Fig:ProbGraph1}, represented by the arrow from $\vect X$ to $Y$, is approximated by $f_{\hat Y|\vect X}$, leading to a new dependency structure as illustrated in Figure \ref{Fig:ProbGraph2}. Consequently, an ``approximate" surrogate model can be expressed by: 
\begin{equation}  \label{Eq:approxmod}
    f_{\hat Y|\vect X}^{(1)}(\hat y|\vect x) = \int f_{\hat Y|\vect\Psi_{\vect z}}(\hat y|\vect\psi_{\vect z})f_{\vect\Psi_{\vect z}|\vect XY}(\vect\psi_{\vect z}|\vect x y')f_{\hat Y|\vect X}^{(0)}(y'|\vect x)\,d\vect\psi_{\vect z}\,dy' \,,
\end{equation}
where $f_{\hat Y|\vect X}^{(0)}\equiv f_{\hat Y|\vect X}$ is the true surrogate model from Eq.~\eqref{Eq:exactmod}, and $f_{\hat Y|\vect X}^{(1)}$ is an approximate surrogate model after one iteration of approximation. This equation does not resolve the ``Catch-22" dilemma; however, if we presume that the transition kernel $f_{\hat Y|\vect\Psi_{\vect z}}\cdot f_{\vect\Psi_{\vect z}|\vect XY}$ encodes a stationary distribution, iterations of Eq.~\eqref{Eq:approxmod} would lead to a fixed-point equation:
\begin{equation}  \label{Eq:approxmod1}
    f_{\hat Y|\vect X}^{(\infty)}(\hat y|\vect x) = \int f_{\hat Y|\vect\Psi_{\vect z}}(\hat y|\vect\psi_{\vect z})f_{\vect\Psi_{\vect z}|\vect XY}(\vect\psi_{\vect z}|\vect x y')f_{\hat Y|\vect X}^{(\infty)}(y'|\vect x)\,d\vect\psi_{\vect z}\,dy' \,,
\end{equation}
which corresponds to the stationarity equation for a Markov process $\{\hat{Y}^{(t)}|\vect X=\vect x\}$ with the transition kernel: 
\begin{equation}  \label{Eq:transition}
   T\left(\hat{y}^{(t)},\hat{y}^{(t+1)}|\vect x\right) = f_{\hat Y|\vect\Psi_{\vect z}}\left(\hat{y}^{(t+1)}|\vect\psi_{\vect z}\right)f_{\vect\Psi_{\vect z}|\vect{X} Y}\left(\vect\psi_{\vect z}|\vect x \hat{y}^{(t)}\right) \,. 
\end{equation}
Accordingly, a random sequence generated from this kernel can be interpreted as samples drawn from $f_{\hat Y|\vect X}^{(\infty)}$. This constitutes the core of the proposed method. 

The transition kernel is derived from an implicit fixed-point formulation of the input–output dimensionality reduction and the predictive distribution. By repeatedly propagating a candidate output through the feature space conditional distribution and the dimensionality reduction mapping, the kernel is designed to converge toward a stationary distribution that reflects the surrogate behavior of the original computational model. This builds on the principle that high-dimensional joint distributions of inputs and outputs can often be represented in a low-dimensional embedding, provided that the dimensionality reduction adequately preserves the essential features of the input–output relationship.

The fundamental assumption of the above approximation scheme is that the transition kernel admits a stationary distribution. If the computational model is deterministic and the dimensionality reduction and feature space conditional distribution are sufficiently accurate, the transition kernel would resemble the Dirac delta $\delta(y-\mathcal{M}(\vect x))$, and the distribution that satisfies the detailed balance condition is also $\delta(y-\mathcal{M}(\vect x))$. Therefore, the approximation scheme is correct at least in the idealized case. Section \ref{Exact_example} offers numerical validation of the approximation scheme under well-controlled conditions (where a perfect dimensionality reduction can be formulated), whereas Section \ref{bar_example} and Section \ref{BW_example} examine more realistic examples. Interestingly, the approximation scheme finds a parallel in statistical linearization techniques \cite{caughey1963equivalent,roberts2003random,elishakoff2017sixty} from nonlinear stochastic dynamics, where determining the optimal surrogate model (equivalent linear system) requires knowing quantities of the original model. To address this dilemma, statistical linearization involves replacing the quantities of the original model with solutions from the to-be-determined surrogate model. This substitution initiates a fixed-point iteration process to identify the surrogate model. 

\begin{figure}[H]
  \centering
  \includegraphics[scale=0.55]  {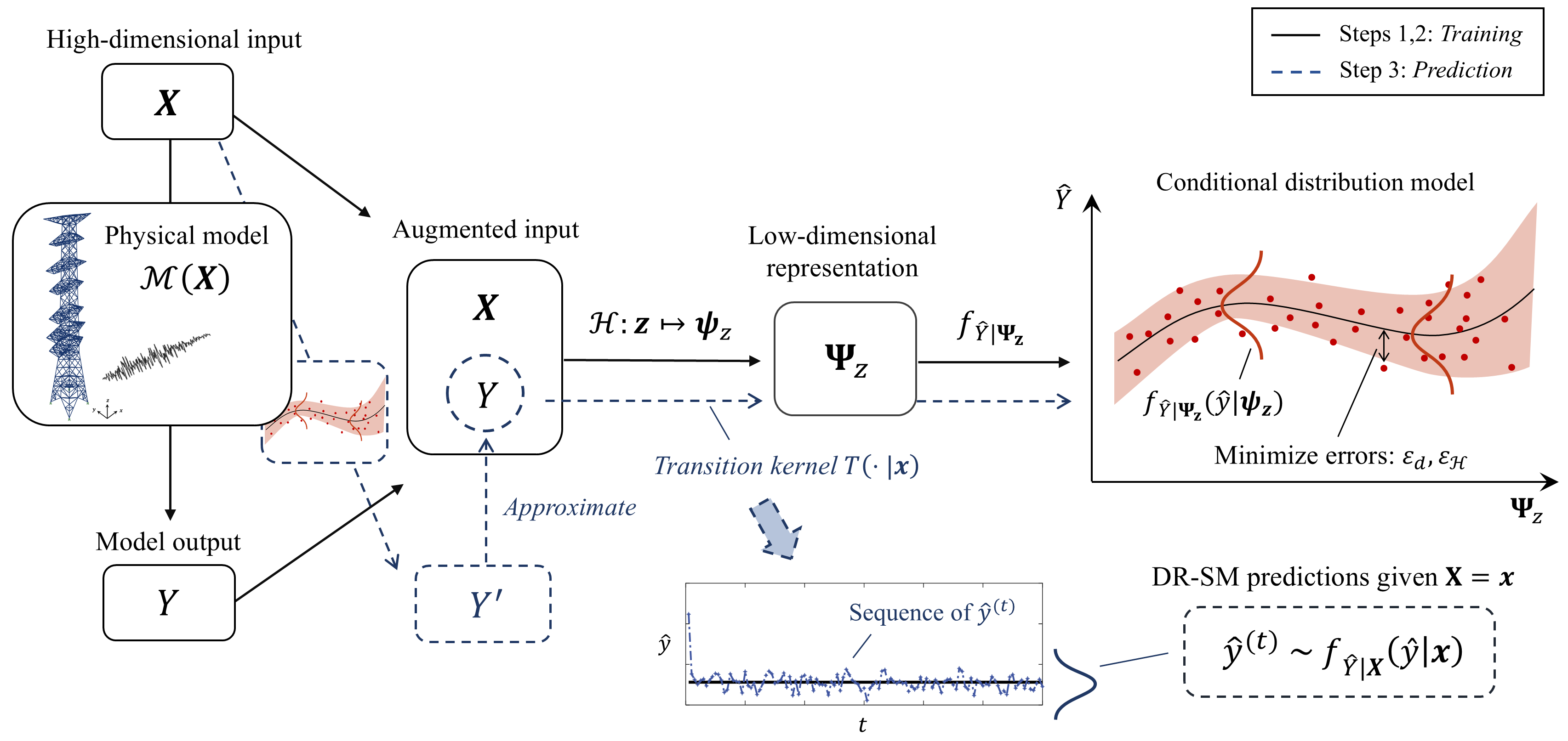}
  \caption{\textbf{Illustration of Dimensionality Reduction-based Surrogate Modeling.} The solid arrows represent the training stage, where the dimensionality reduction map $\mathcal{H}$ and the feature space conditional distribution $f_{\hat Y|\vect\Psi_{\vect z}}$ are trained. The dashed arrows denote the prediction stage, which utilizes the trained $\mathcal{H}$ and $f_{\hat Y|\vect\Psi_{\vect z}}$ as transition kernels to generate samples from $f_{\hat Y|\vect X}(\hat y|\vect X=\vect x)$, thereby predicting $y$ given $\vect x$.}
  \label{Fig:Illustration}
\end{figure}

To conclude this section, Figure \ref{Fig:Illustration} summarizes the proposed surrogate modeling method. 

\section{Computational details of DR-SM}  \label{DRSMdetails}
\subsection{Dimensionality reduction} \label{DRmapping}
We presume that the hyper-surface $\{(\vect{x},y)\in\mathbb{R}^{n+1}: y-\mathcal{M}(\vect{x})=0\}$ admits a lower-dimensional representation even when the input random variables $\vect X$ are genuinely high-dimensional. Provided with a training set $\{\vect z^{(i)}\}_{i=1}^N\equiv\{(\vect{x}^{(i)},y^{(i)})\}_{i=1}^N$, where $y^{(i)}\equiv\mathcal{M}(\vect{x}^{(i)})$, we construct a dimensionality reduction mapping $\mathcal{H}: \vect z\in\r^{n+1}\mapsto\vect\psi_{\vect z}\in\r^d$. The proposed DR-SM is not confined to any specific dimensionality reduction technique, while the principle to select the reduced dimension $d$ can be standardized by evaluating the accuracy of the feature space conditional distribution model. 

\subsection{Conditional distribution model} \label{hGPmodel}
Given the dimensionality reduction mapping $\mathcal{H}$ and the training set $\{\vect z^{(i)}\}_{i=1}^N$, the set $\{(\vect{\psi}_{\vect z}^{(i)},y^{(i)})\}_{i=1}^N$ can be obtained. Subsequently, we construct a conditional distribution model $f_{\hat Y|\vect{\vect\Psi_{\vect z}}}$ to predict $y$ given a feature vector. In this paper, we adopt the heteroscedastic Gaussian process (hGP) to model $f_{\hat Y|\vect{\vect\Psi_{\vect z}}}$. This selection is particularly relevant because of the inherent errors in feature space representation. Therefore, a model capable of accommodating noisy observations is preferred \cite{kim2024adaptive,kim2023estimation}. The prediction of the hGP model given $\vect{\vect\Psi}_{\vect z}=\vect{\vect\psi}_{\vect{z}_*}$ is expressed as: \cite{lazaro2011variational}
\begin{equation}  \label{Eq:hGP_distn}
\hat Y({\vect\psi}_{\vect{z}_*}) \sim \mathcal{N}\left(\mu_{\hat{Y}|{\vect\Psi}_{\vect{z}}}({\vect\psi}_{\vect{z}_*}), \sigma_{\hat{Y}|{\vect\Psi}_{\vect{z}}}^2({\vect\psi}_{\vect{z}_*}) \right) \,, \,\,\,
\end{equation}
where $\mu_{\hat{Y}|{\vect\Psi}_{\vect{z}}}({\vect\psi}_{\vect{z}_*})$ and $\sigma_{\hat{Y}|{\vect\Psi}_{\vect{z}}}^2({\vect\psi}_{\vect{z}_*})$ are respectively the conditional mean and variance. Details of the hGP model inference are given in \ref{App:hGP}. It is noted that the hGP model hyperparameters are calibrated once during the initial training stage, and remain fixed throughout the subsequent stochastic sampling procedure. The hGP model is selected in this paper due to its capacity to model heteroscedastic noise, which arises from compression errors in dimensionality reduction. However, the proposed framework is not restricted to this model and can readily accommodate alternative choices, such as KDE \cite{jia2017new} or conditional deep surrogates \cite{yang2019conditional}.

\subsection{Optimizing the dimensionality reduction parameters} \label{Training}
While DR-SM allows for a flexible choice of dimensionality reduction method, the selection of its parameters can be tailored to minimize the surrogate prediction error. The proposed DR-SM takes a two-step approach, initially estimating the optimal dimension $d$ and subsequently fine-tuning the parameters $\vect{\theta}_{\mathcal{H}}$. To this end, Algorithm \ref{alg:dimension} \cite{kim2024adaptive} is designed to find $d$ that yields acceptable mean squared error:
\begin{equation}  \label{Eq:error_d}
\varepsilon_d = \sqrt{\frac{1}{N}\sum_{i=1}^{N}\left(y^{(i)}-\mu_{\hat{Y}|{\vect\Psi}_{\vect{z}}}\left({\vect\psi}_{\vect{z}}^{(i)}; d, \vect{\theta}_{\mathcal{H}} \right)\right)^2} \,,
\end{equation}
where $\mu_{\hat{Y}|{\vect\Psi}}$ is the mean prediction from the hGP model. The error $\varepsilon_d$ quantifies the impact of $d$ on the accuracy of the feature space conditional distribution model. The fixed value of the parameter $\vect{\theta}_{\mathcal{H}}$ is used during this procedure. Starting from $d=1$, $d$ is iteratively increased until $\varepsilon_d$ becomes no greater than a specified threshold $\varepsilon_d^t$.

\begin{breakablealgorithm}
\label{alg:dimension}
\caption{Determine the optimal reduced dimension $d^{*}$.}
\begin{description}
Given a set of training data $\{\vect x^{(i)},y^{(i)}\}_{i=1}^{N}$:
\\
$d\leftarrow{0}$; $\varepsilon_d\leftarrow\infty$;\\
While $\varepsilon_d>\varepsilon_d^t$, do
\\
\,\,\,\,\,\,\,$d\leftarrow{d+1}$; 
\\
\,\,\,\,\,\,\,Identify the $d$-dimensional feature mapping parameterized by $\vect{\theta}_{\mathcal{H}}$;
\\
\,\,\,\,\,\,\,Compute the mean predictions $\{\mu_{\hat{Y}|{\vect\Psi}_{\vect{z}}}({\vect\psi}_{\vect{z}}^{(i)}; d,\vect{\theta}_{\mathcal{H}})\}_{i=1}^N$;
\\
\,\,\,\,\,\,\,Compute $\varepsilon_d$;
\\
End
\\
$d^*\leftarrow{d}$; 
\end{description}
\end{breakablealgorithm}
Next, the dimensionality reduction parameters $\vect{\theta}_{\mathcal{H}}$ can be fine-tuned using the following optimization:
\begin{equation} \label{Eq:optim_MLpara}
\vect{\theta}^*_{\mathcal{H}}=\mathop{\arg\min}_{\vect{\theta}_{\mathcal{H}}\in \vect{\Omega}_{\vect{\theta}_{\mathcal{H}}}} \varepsilon_{d^*}(\vect{\theta}_{\mathcal{H}}) \,,
\end{equation}
where $\vect{\Omega}_{\vect{\theta}_{\mathcal{H}}}$ is the admissible set for parameters of the selected dimensionality reduction method. It is noted that identifying $d^*$ and fine-tuning $\vect{\theta}_{\mathcal{H}}$ are sequentially performed using a fixed training set, thereby eliminating the need for additional evaluations of the model $\mathcal{M}$.

It is noted that the performance of the proposed method is sensitive to both the selection of the reduced dimension and the number of training samples. While the reduced dimension is chosen based on a feature space error threshold, overfitting may occur if the training data is insufficient or scarce. Parametric studies in Section \ref{Examples} illustrate this trade-off.

\subsection{Extracting a stochastic surrogate model} \label{Prediction}
Given an input $\vect{X}=\vect{x}$, a random sequence generated by the transition kernel in Eq.~\eqref{Eq:transition} is used to approximate samples of $f_{\hat Y|\vect X}$, collectively representing the stochastic surrogate model prediction. The starting point of the random sequence is set to the average of the training outputs to initialize the random sequence near the center of the output distribution. This sampling procedure is described in Algorithm \ref{alg:Tsampling}. Note that this procedure is a special case of Markov Chain Monte Carlo sampling \cite{gilks1995markov}, with a transition kernel $T(\cdot|\vect x)$ derived directly from the stationary condition. From the random samples, we can estimate the mean $ \mu_{\hat Y}(\vect x) = \mathbb{E}[\hat Y|\vect x ]$, variance $\sigma^2_{\hat Y}(\vect x) = \mathbb{V}\text{ar}[\hat Y|\vect x ]$, and the quantile bounds of the prediction:
\begin{equation}  \label{Eq:StoSurrogate}
\begin{aligned}
y^{+}&=\mu_{\hat Y}(\vect x) + \kappa\sigma_{\hat Y}(\vect x)  \,,   \\   y^{-}&=\mu_{\hat Y}(\vect x) - \kappa\sigma_{\hat Y}(\vect x)  \,,  
\end{aligned}
\end{equation}
where $\kappa > 0$ is a parameter controlling the bounds. This quantified prediction variability can facilitate an integration with active learning \cite{marelli2018active,kim2024adaptive} strategies to further improve the efficiency of surrogate modeling.
\begin{breakablealgorithm}
\label{alg:Tsampling}
\caption{Sampling procedure using the transition kernel.}
\begin{description}
Given the tuned parameters $d^*$ and $\vect{\theta}^*_{\mathcal{H}}$, and test input $\vect{x}$:
\\
Set a starting point $\hat{y}^{(1)}= \frac{1}{N} \sum_{i=1}^{N} y^{(i)}$;
\\
For $t=1,...,N_t-1$
\\
\,\,\,\,\,\,\,Identify the reduced sample point through the feature mapping: $\mathcal{H}:(\vect{x},\hat y ^{(t)}) \mapsto \vect{\psi}_{\vect z}^{(t)}$;
\\
\,\,\,\,\,\,\,Draw a random sample from conditional distribution: $y' \sim f_{\hat Y|\vect\Psi_{\vect z}}\left(\hat{y}|\vect{\Psi}_{\vect z}=\vect{\psi}_{\vect z}^{(t)} \right)$;
\\
\,\,\,\,\,\,\,$\hat y ^{(t+1)} \leftarrow y'$;
\\
End
\\
Exclude the first $N_b$ samples as burn-in and obtain the sequence of samples $\{\hat{y}^{(t)}\}_{t=N_b+1}^{N_t}$; 
\end{description}
\end{breakablealgorithm}

\subsection{Algorithm of DR-SM}  \label{Algorithm}

\noindent The procedures of DR-SM are summarized in Algorithm \ref{alg:DRSM}.
\begin{breakablealgorithm}
\label{alg:DRSM}
\caption{Dimensionality reduction-based Surrogate Modeling.}
\begin{description}
\item [Step 1.] Generate a training set $\mathcal{Z}_\mathcal{D}=\{(\vect{x}^{(i)},\mathcal{M}(\vect x^{(i)}))\}_{i=1}^N$.
~\
\item [Step 2.] Given $\mathcal{Z}_\mathcal{D}$, identify the optimal dimension $d^*$ using Algorithm \ref{alg:dimension} and fine-tune the dimensionality reduction parameters $\vect{\theta}_{\mathcal{H}}^{*}$ by solving Eq.~\eqref{Eq:optim_MLpara}. For each trial value in the optimization procedures, the feature space conditional distribution model is trained simultaneously. This stage yields the feature mapping $f_{\vect\Psi_{\vect z}|\vect{X} Y}$ and the feature space conditional distribution model $f_{\hat Y|\vect\Psi_{\vect z}}$, which constitute the transition kernel expressed by Eq.~\eqref{Eq:transition}.

\item [Step 3.] Given an input $\vect{x}$, generate a sequence of random samples from the transition kernel $T(\hat{y}^{(t)},\hat{y}^{(t+1)}|\vect x)$, as described in Algorithm \ref{alg:Tsampling}. The obtained samples $\{\hat{y}^{(t)}\}_{t=N_b+1}^{N_t}$ are used as the stochastic surrogate model predictions at $\vect x$. 
\end{description}
\end{breakablealgorithm}

\section{Numerical examples} \label{Examples}
This section investigates three high-dimensional UQ problems to demonstrate the performance of DR-SM. First, the method is applied to analytical linear model to investigate numerical validation of the approximation scheme in DR-SM, in which a perfect dimensionality reduction can be formulated. Next, the method is applied to more realistic engineering examples: (i) a linear elastic bar whose axial rigidity is modeled by a random field and (ii) a hysteretic structural system subjected to random process excitation. For these examples, three dimensionality reduction techniques are adopted: PCA, kernel-PCA (with polynomial kernel), and autoencoder. The training sets of $\vect X$ are generated by Latin Hypercube Sampling with sample decorrelation, and the reference solutions are obtained from the direct MCS. The total number of fixed-point iterations and the burn-in length are set to $N_t=1200$ and $N_b=200$, respectively.

Through the examples, the following relative mean squared error $\varepsilon_{\eta}$ is employed to investigate the accuracy of DR-SM:
\begin{equation} \label{Eq:error_model}
\varepsilon_{\eta} =\frac{\mathbb{E}\left[\left(\mu_{\hat Y}(\vect{X}) - \mathcal{M}(\vect{X})\right)^{2}\right]}{\mathbb{V}\text{ar}\left[\mathcal{M}(\vect{X})\right]}\,,
\end{equation}
\noindent where $\mu_{\hat Y}$ is the mean prediction of DR-SM.

\subsection{Illustration example: High-dimensional linear function} \label{Exact_example}


\noindent First, a linear function is considered to investigate the performance of DR-SM in the case of perfect dimensionality reduction. The model is formulated as
\begin{equation} \label{Eq:Exactlinear}
Y = \mathcal{M}(\vect{X}) = \beta_0 - \sum_{i=1}^{n}{X_i} \,, 
\end{equation}
\noindent where $X_i, \, i=1,2,...,n$, are standard Gaussian random variables, $n$ is the dimension, and $\beta_0$ is a known model parameter. This is a special model that allows for a perfect dimensionality reduction through a linear transformation \cite{kim2024adaptive}. The perfect dimensionality reduction is achieved when the low-dimensional representation does not induce a loss of information in predicting the quantities of interest. One such transformation is
\begin{equation} \label{Eq:Exactfeature}
\Psi_z = \sum_{i=1}^{n}{X_i}+\frac{1}{10}(Y-\beta_0) = \frac{9}{10}\sum_{i=1}^{n}{X_i} \,, 
\end{equation}
\noindent from which an exact mapping to the output exists:
\begin{equation} \label{Eq:ExactMapping}
Y = \beta_0-\frac{10}{9}\Psi_z\,.
\end{equation}
\noindent With Eqs.~\eqref{Eq:Exactlinear} and~\eqref{Eq:Exactfeature}, the first two terms in the integrand of Eq.~\eqref{Eq:exactmod} become Dirac delta functions, i.e., $f_{\Psi_{z}|\vect XY}(\psi_z|\vect x,y)=\delta(\psi_z-\sum_{i=1}^{n}{x_i}-0.1(y-\beta_0))$ and $f_{\hat Y|\Psi_{\vect z}}(\hat y|\psi_{z}) = \delta(\hat y - \beta_0 + 1/0.9\psi_z)$, respectively. Consequently, $f_{\hat Y|\vect X}$ equals to $f_{Y|\vect X}$. This makes Eq.~\eqref{Eq:approxmod} an exact equivalent of Eq.~\eqref{Eq:exactmod} and, as a result, the stationary distribution encoded in the transition kernel is the exact response.

\begin{figure}[H]
  \centering
  \includegraphics[scale=0.48] {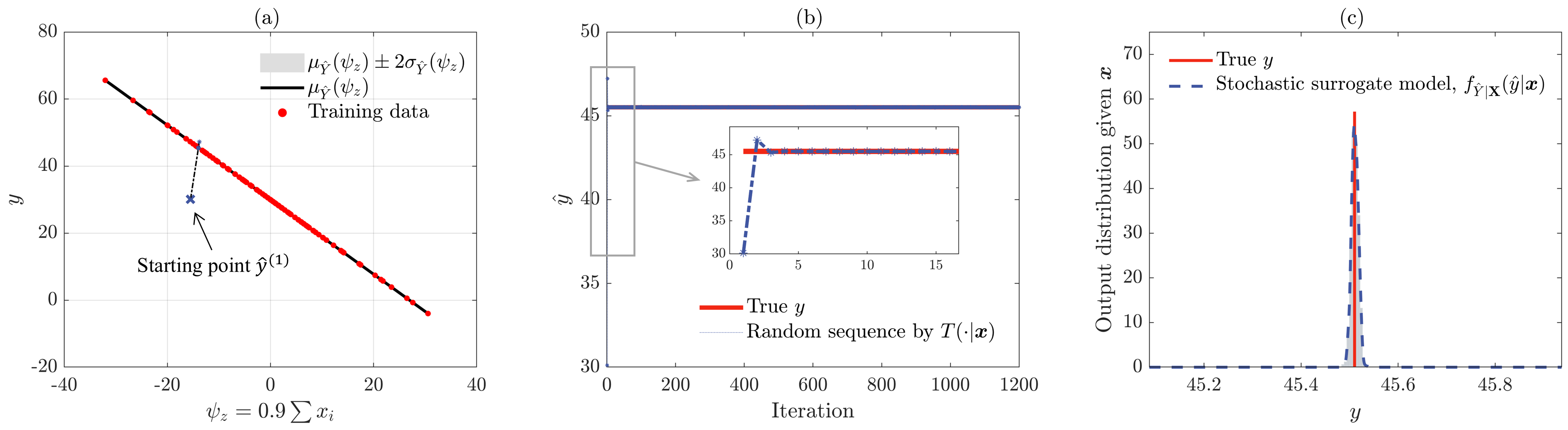}
  \caption{\textbf{DR-SM for the high-dimensional linear example: (a) the ``exact” feature mapping, (b) trajectories of $\hat y$, and (c) stochastic surrogate model.} This figure is obtained using the DR-SM equipped with the ``exact" dimensionality reduction. In figure (b), random samples $\hat y ^{(t)}$ are generated by the transition kernel $T(\hat y^{(t)},\hat y^{(t+1)}| \vect x)$. In figure (c), stochastic surrogate model $f_{\hat Y|\vect X}$ is obtained using the sequence of random samples $\hat y ^{(t)}$ presented in figure (b), and red solid line represents the true value of $y$ given $\vect x$.}
  \label{Fig_A_exact}
\end{figure}

For a numerical demonstration, Algorithm \ref{alg:DRSM} is applied to this linear model. To enforce the perfect dimensionality reduction, Eq.~\eqref{Eq:Exactfeature} is directly used, dropping the parameter optimization step. The mapping in Eq.~\eqref{Eq:ExactMapping} is considered unknown and is trained using hGP. Due to the limited number of training samples, the surrogate prediction is expected to be a narrow distribution rather than a deterministic value. Figure \ref{Fig_A_exact} shows the stochastic surrogate modeling procedure using 100 training points for an arbitrarily selected test input $\vect X = \vect x$. The model parameter is $\beta_0=30$ and the input dimension is $n=100$. Starting from the initial value $\hat{y}^{(1)}=30$, the random samples are generated by the transition kernel $T(\hat y^{(t)},\hat y^{(t+1)}| \vect x)$ expressed by Eq.~\eqref{Eq:transition}. The results show that the proposed surrogate model quickly converges to the true solution. The stochastic surrogate model $f_{\hat Y|\vect X}(\hat y|\vect x)$ is concentrated around the true response value, as shown in Figure \ref{Fig_A_exact}(c).

\begin{figure}[H]
  \centering
  \includegraphics[scale=0.55] {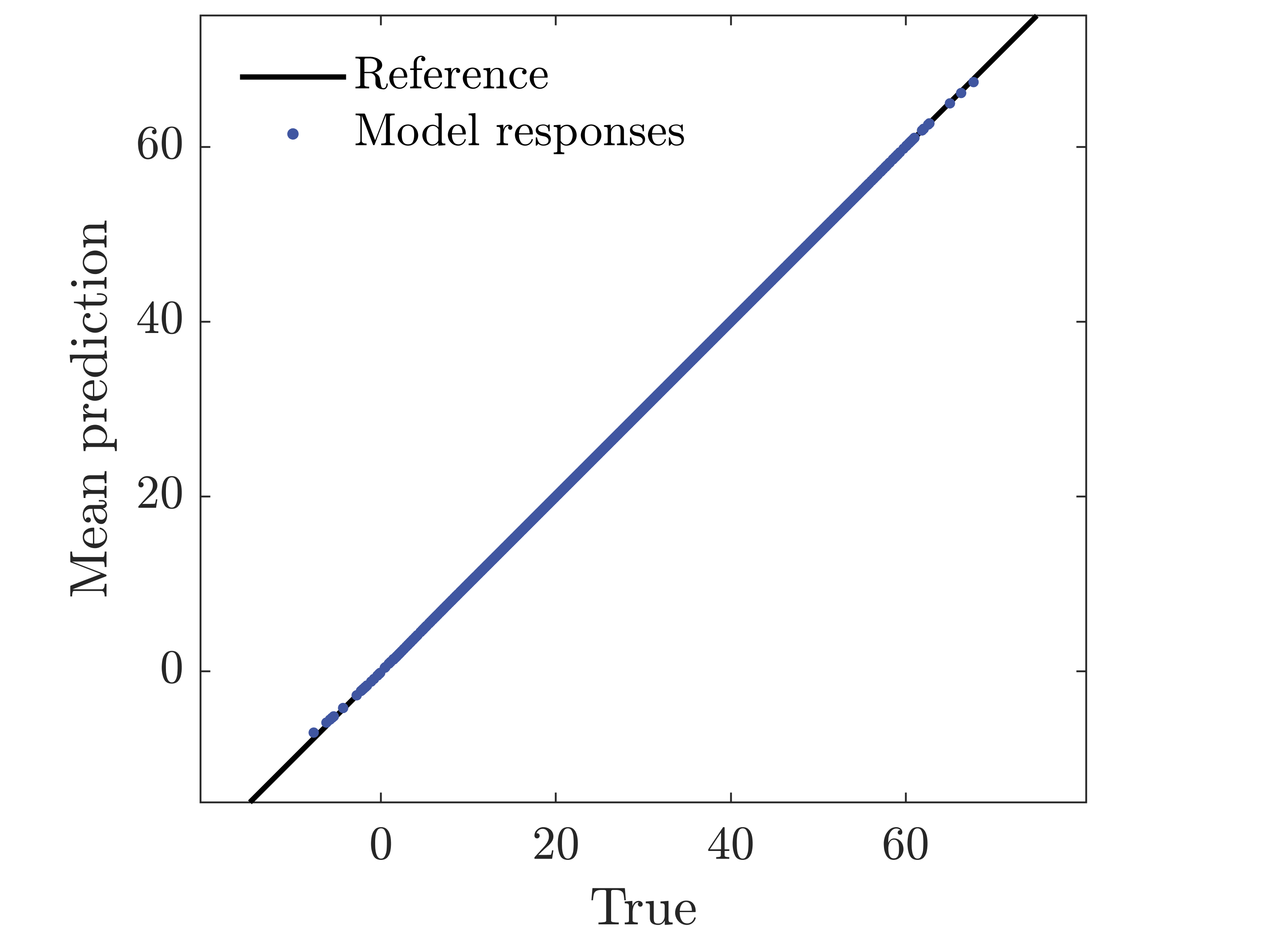}
  \caption{\textbf{A scatter plot of the true model outputs against the surrogate model predictions for the high-dimensional linear example.} The relative mean squared error $\varepsilon_{\eta}$ is calculated as $1.3961\times10^{-6}$.}
  \label{Fig_A_exactYY}
\end{figure}

Figure \ref{Fig_A_exactYY} shows the scatter plot of the true model responses against the surrogate predictions $\mu_{\hat Y}$, obtained from $10^4$ test points. In the ideal case of perfect prediction, the scatter plot would be concentrated along the diagonal line. The results demonstrate the near-perfect accuracy of DR-SM when the perfect dimension reduction is achieved.


Next, to further test the performance of DR-SM, the following model with additive noise $\varepsilon$ is considered:
\begin{equation} \label{Eq:Exactnoise}
Y = \mathcal{M}(\vect{X}) = \beta_0 - \sum_{i=1}^{n}{X_i} + \varepsilon \,, 
\end{equation}
\noindent where $\varepsilon \sim N(0,1)$. Thus, the true response at $\vect x$ is represented by a Gaussian distribution $\mathcal{N}(\beta_0 - \sum_{i=1}^{n}{x_i}, 1)$. The pefect dimensionality reduction is applied, i.e., $f_{\Psi_{z}|\vect XY}(\psi_z|\vect x,y)=\delta(\psi_z-\sum_{i=1}^{n}{x_i}-0.1(y-\beta_0))$. 

\begin{figure}[H]
  \centering
  \includegraphics[scale=0.48] {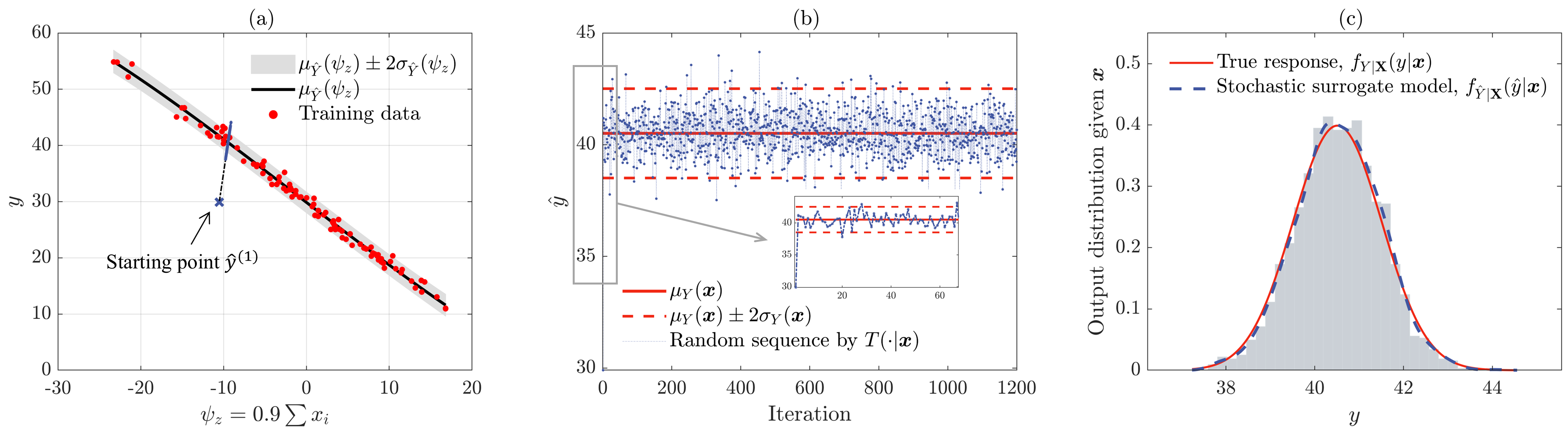}
  \caption{\textbf{DR-SM for the high-dimensional linear example with additive noise: (a) the ``exact” feature mapping, (b) trajectories of $\hat y$, and (c) stochastic surrogate model.} This figure is obtained using the DR-SM equipped with the ``exact" dimensionality reduction. In figure (c), stochastic surrogate model $f_{\hat Y|\vect X}$ is obtained using the sequence of random samples $\hat y ^{(t)}$ presented in figure (b), and red solid line represents the true probabilistic response $f_{Y|\vect X}$ given $\vect x$.}
  \label{Fig_A_exactnoise}
\end{figure}

\begin{figure}[H]
  \centering
  \includegraphics[scale=0.54] {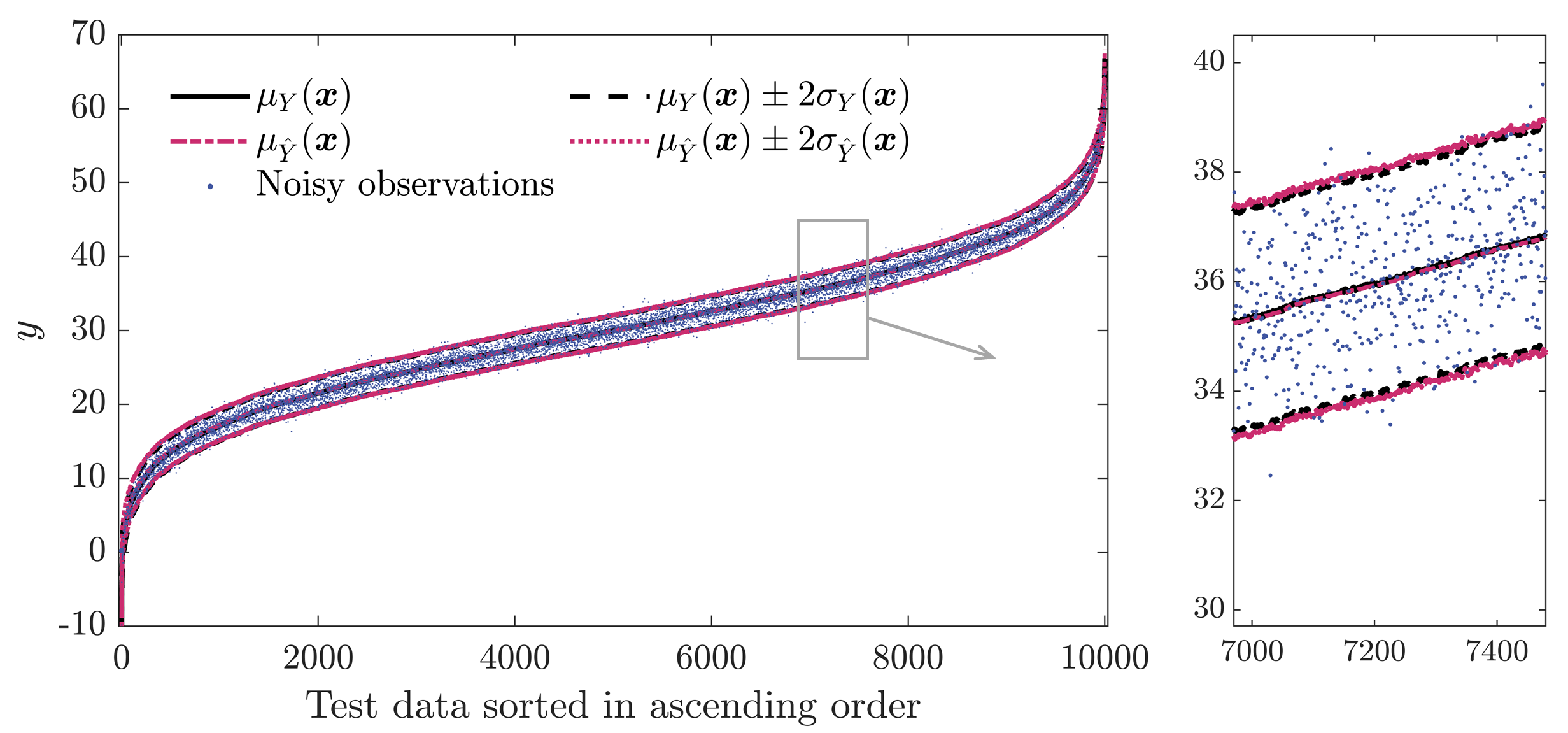}
  \caption{\textbf{A scatter plot of the true model outputs against the surrogate model predictions with uncertainty intervals for the high-dimensional linear example with additive noise.} The noisy observations are represented by blue circles. Surrogate-based mean predictions and the uncertainty intervals are represented as red dashed and dotted lines, respectively, while true mean and the uncertainty intervals are represented as black solid and dashed lines.}
  \label{Fig_A_exactnoiseYY}
\end{figure}

Figure \ref{Fig_A_exactnoise} presents the stochastic surrogate modeling procedure using 100 training points for an arbitrarily selected test input $\vect X = \vect x$. Due to the additive noise, the true response at $\vect x$ is probabilistic, as shown in Figure \ref{Fig_A_exactnoise}(c). Figure \ref{Fig_A_exactnoiseYY} presents the predicted mean and prediction intervals for $10^4$ randomly generated test points. The test samples are rearranged in the ascending order of the predicted mean. The surrogate-based predicted means and standard deviation intervals (black solid and dashed lines) are compared with those obtained by true model (red dashed and dotted lines). Noisy observation are denoted as blue circles. A close-up is provided to visualize the details of the performance. Since the prediction is probabilistic, an uncertainty measure can be readily obtained without additional statistical techniques, such as the bootstrap resampling. The results confirm the stochastic surrogate modeling accuracy of DR-SM. 


\subsection{Linear elastic bar with random axial rigidity} \label{bar_example}

\begin{figure}[H]
  \centering
  \includegraphics[scale=0.68] {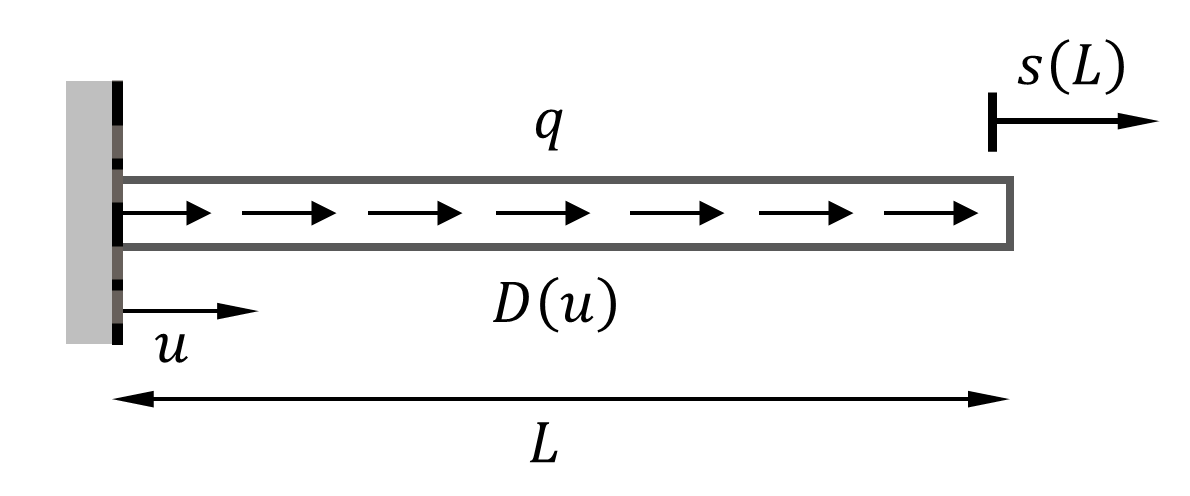}
  \caption{\textbf{Linear elastic bar with random axial rigidity.}}
  \label{Fig_bar}
\end{figure}

\noindent As an engineering application, this example considers a uniaxial linear elastic bar subjected to a uniformly distributed tension load \cite{papaioannou2019pls} in Figure \ref{Fig_bar}. The displacement field $s(u)$ is governed by the following differential equation:
\begin{equation} \label{Eq:bar_eq}
\frac{d}{du}\left(D(u)\frac{d}{du}s(u)\right) + q(u) = 0\,, \,\,\,0<u<L\,,
\end{equation}
\noindent where $L$ is the length of the bar; and $D(u)=AE(u)$ is the axial resistance of the bar, which is described by a homogeneous random field in one spatial dimension. $D$ has a lognormal marginal distribution with mean $\mu_{D}=100$ kN and standard deviation $\sigma_{D}=10$ kN. The autocorrelation function of the underlying Gaussian random field $\ln{D}$ is modeled by an isotropic exponential function as
\begin{equation} \label{Eq:bar_autocor}
\rho_{\ln{D}}(\Delta{u}) = \exp{(-|\Delta{u}|/\gamma)}\,,
\end{equation}
\noindent where $\gamma$ is the correlation length set to $\gamma=0.04$ m. The random field $\ln{D}$ is discretized by a Karhunen-Loève (KL) expansion \cite{ghanem2003stochastic,betz2014numerical}, i.e., 
\begin{equation} \label{Eq:bar_KL}
D(u) = \exp{\left\lbrace \mu_{\ln{D}} + \sigma_{\ln{D}}\sum_{i=1}^{N_{KL}}\sqrt{\lambda_{i}}\varphi_{i}(u)\xi_{i} \right\rbrace}\,,
\end{equation}
\noindent where $\mu_{\ln{D}}$ and $\sigma_{\ln{D}}$ are distribution parameters of the lognormal variable $D$; $\lambda_{i}$ and $\varphi_{i}(u)$ respectively denote the eigenvalues and eigenfunctions of the correlation kernel in Eq.~\eqref{Eq:bar_autocor}; and $\xi_{i}$ are standard Gaussian random variables, denoted as KL terms. The number of KL terms is $N_{KL}=100$, which captures 95\% of the variability of $\ln{D}$. Therefore, the input vector consists of 100 independent standard Gaussian random variables, i.e., $\vect{X}=[\xi_{1},...,\xi_{100}]$. The bar is subjected to a deterministic load $q =1$ kN/m and the bar length is $L=1$ m. Eq.~\eqref{Eq:bar_eq} is solved by the finite element (FE) method with 100 piecewise linear elements. The response quantity of interest is the displacement at $u=L$ (tip of the bar), i.e., $Y = \mathcal{M}(\vect X) = s(L;\vect X)$.

\begin{figure}[H]
  \centering
  \includegraphics[scale=0.48] {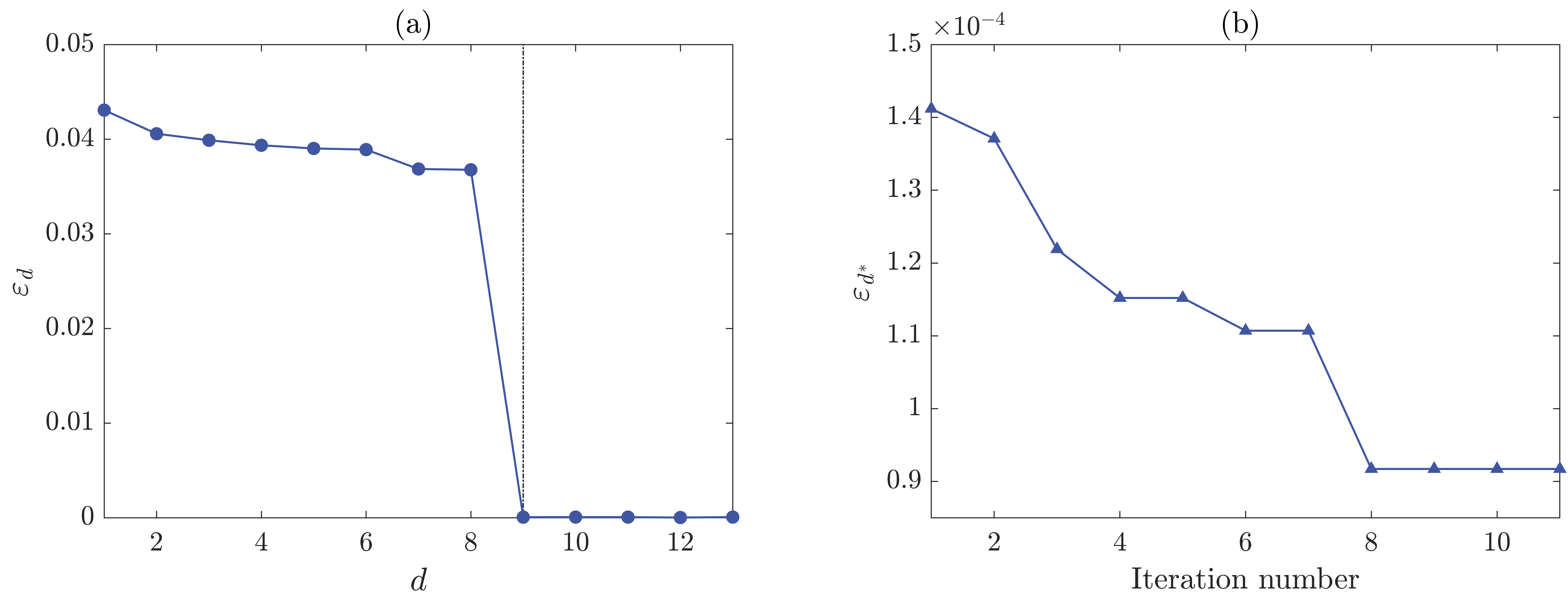}
  \caption{\textbf{(a) Error $\varepsilon_d$ as a function of reduced dimension and (b) error $\varepsilon_{d^*}$ during the fine-tuning of the dimensionality reduction parameters for the linear elastic bar example.} The figure (a) is a byproduct of Algorithm 1, suggesting $d=9$ is sufficient.}
  \label{Figure_bar_d_para}
\end{figure}

Using a training set of 600 samples, kernel-PCA \cite{van2009dimensionality,kontolati2022survey} with the polynomial kernel $\kappa = (\vect{z}\vect{z}^T+a)^b$ is first applied to construct DR-SM. The parameters are $\vect{\theta}_{\mathcal{H}}=[a,b]$, where $a\in\r_{\geq 0}$, $b\in\mathbb{Z}_{\geq 0}$. The threshold for reduced dimension in Algorithm \ref{alg:dimension}, $\varepsilon_d^t$, is set to 0.001. Figure \ref{Figure_bar_d_para}(a) illustrates the change in the mean squared error, described by Eq.~\eqref{Eq:error_d}, as the reduced dimension varies. The result shows that $d=9$ is sufficient to achieve a small prediction error. Figure \ref{Figure_bar_d_para}(b) shows the history of the error during the optimization process of Eq.~\eqref{Eq:optim_MLpara}. Through this process, the parameters for kernel-PCA are fine-tuned to $\vect{\theta}^*_{\mathcal{H}} =[1.236,2]$. 

\begin{figure}[H]
  \centering
  \includegraphics[scale=0.65] {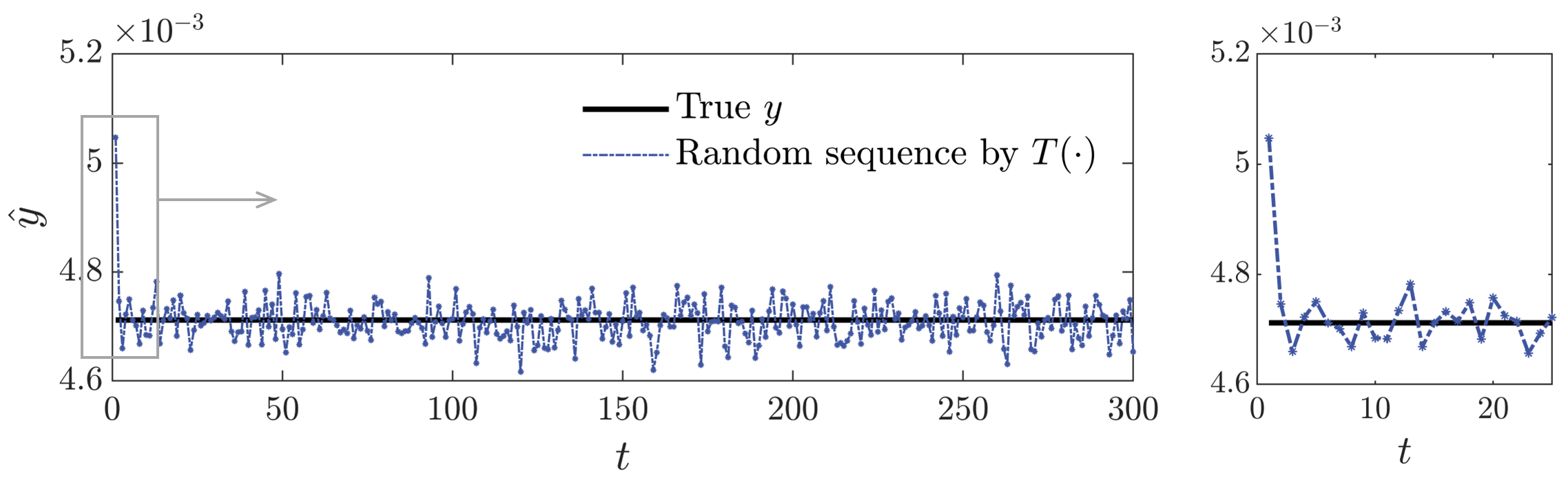}
  \caption{\textbf{Trajectories of $\hat y$ obtained from DR-SM for linear elastic bar example.} Random samples $\hat y ^{(t)}$ are generated by the transition kernel $T(\hat y^{(t)},\hat y^{(t+1)}| \vect x)$. Solid line represents the true value of $y$ given $\vect x$.}
  \label{Fig_sequence_bar}
\end{figure}

\begin{figure}[H]
  \centering
  \includegraphics[scale=0.55] {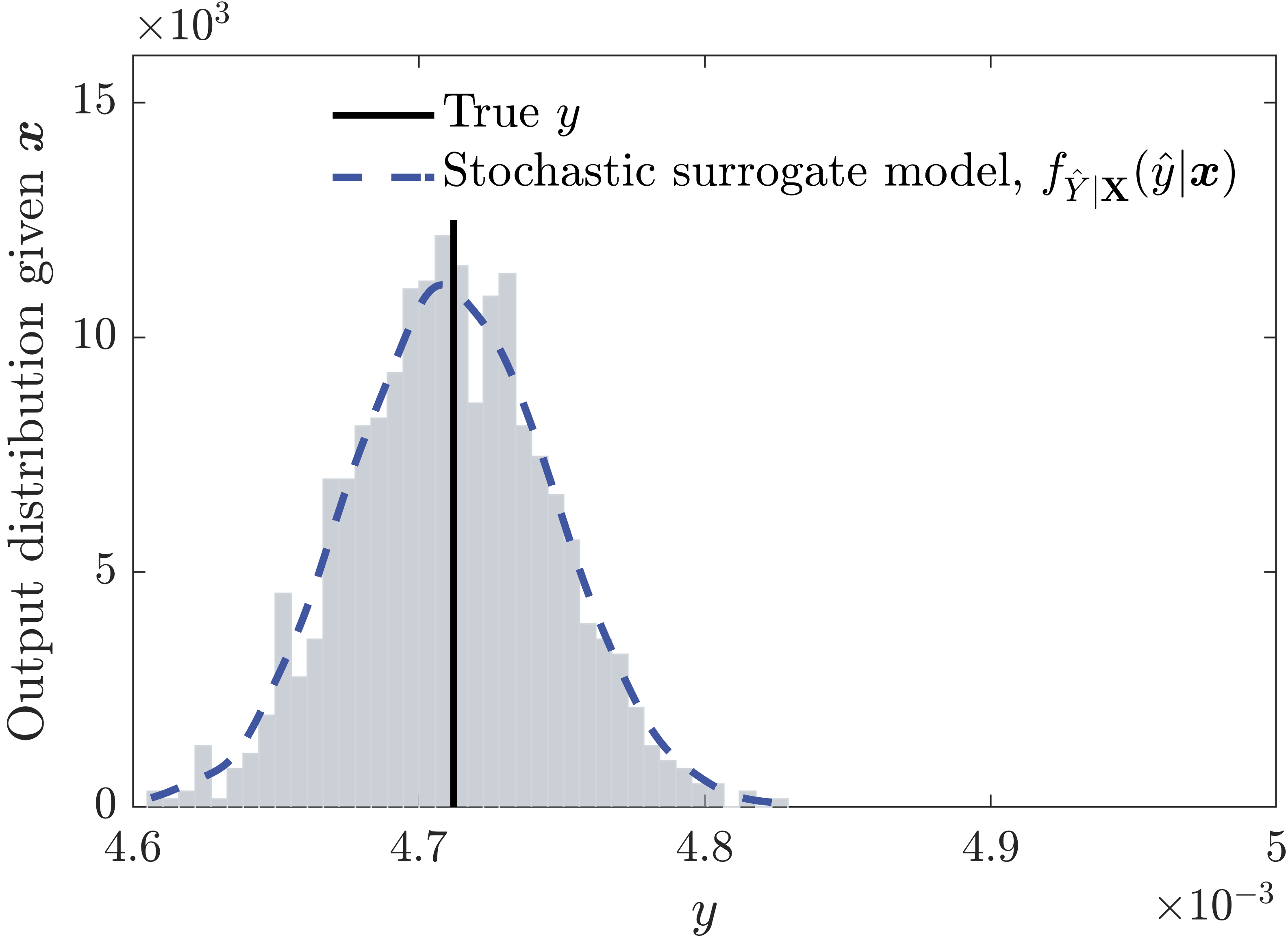}
  \caption{\textbf{Stochastic surrogate model extracted from the dimensionality reduction for linear elastic bar example.} Stochastic surrogate model $f_{\hat Y|\vect X}$ is obtained using the sequence of random samples $\hat y ^{(t)}$ presented in Figure \ref{Fig_sequence_bar}. Solid line represents the true value of $y$ given $\vect x$.}
  \label{Fig_stosurrogate_bar}
\end{figure}

Next, we use DR-SM to make predictions. Figure \ref{Fig_sequence_bar} illustrates the trajectories of the first 300 iterations of the random sequence $\hat y^{(t)}$, $t=1,...,300$, generated for an arbitrarily selected input $\vect x$ using the transition kernel $T(\hat y^{(t)},\hat y^{(t+1)}| \vect x)$ defined in Eq.~\eqref{Eq:transition}. A close-up view is provided to represent the rapid convergence observed during the initial iterations. Figure \ref{Fig_stosurrogate_bar} compares the surrogate prediction with the true $y$. The results show that the surrogate prediction agrees well with the true value.


\begin{figure}[H]
  \centering
  \includegraphics[scale=0.46] {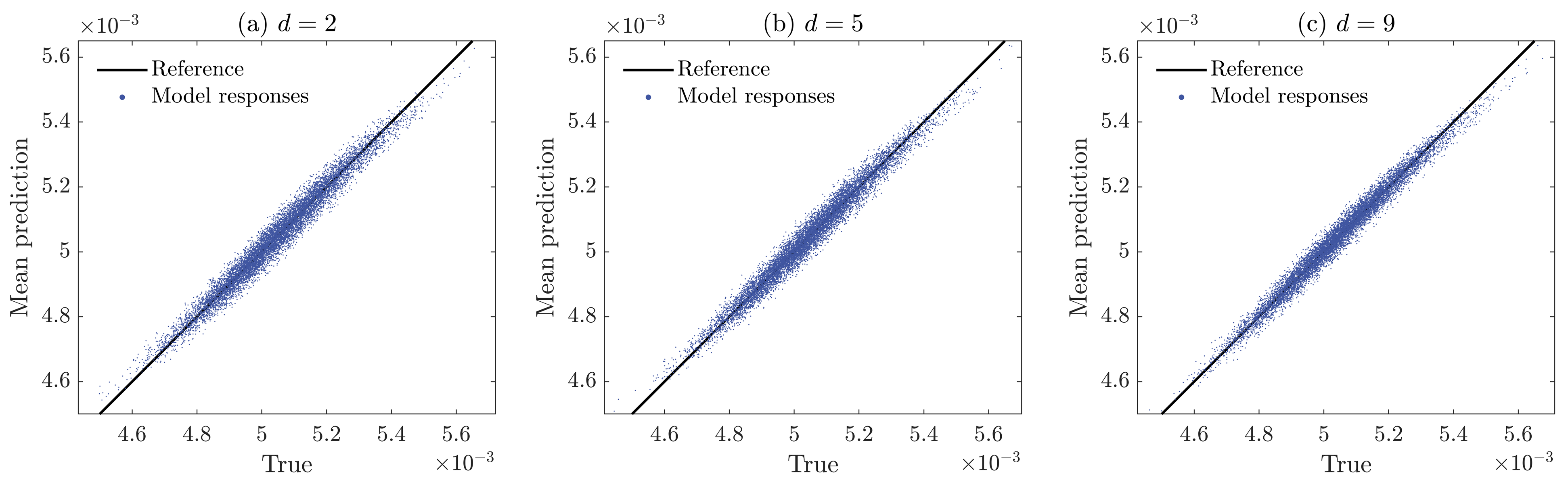}
  \caption{\textbf{A scatter plot of the true model outputs against the surrogate model predictions for the linear elastic bar example with the reduced dimensionality of (a) $d=2$, (b) $d=5$, and (c) $d=9$.} The surrogate model is trained using $600$ samples, the scatter plot is obtained using a test set of $10^4$ samples. The relative mean squared errors $\varepsilon_{\eta}$ are calculated as $0.0325$, $0.0289$, and $0.0220$, respectively. The accuracy is high with the optimal reduced dimension $d=9$.}
  \label{Fig_YYscatter_bar}
\end{figure}

\begin{figure}[H]
  \centering
  \includegraphics[scale=0.58] {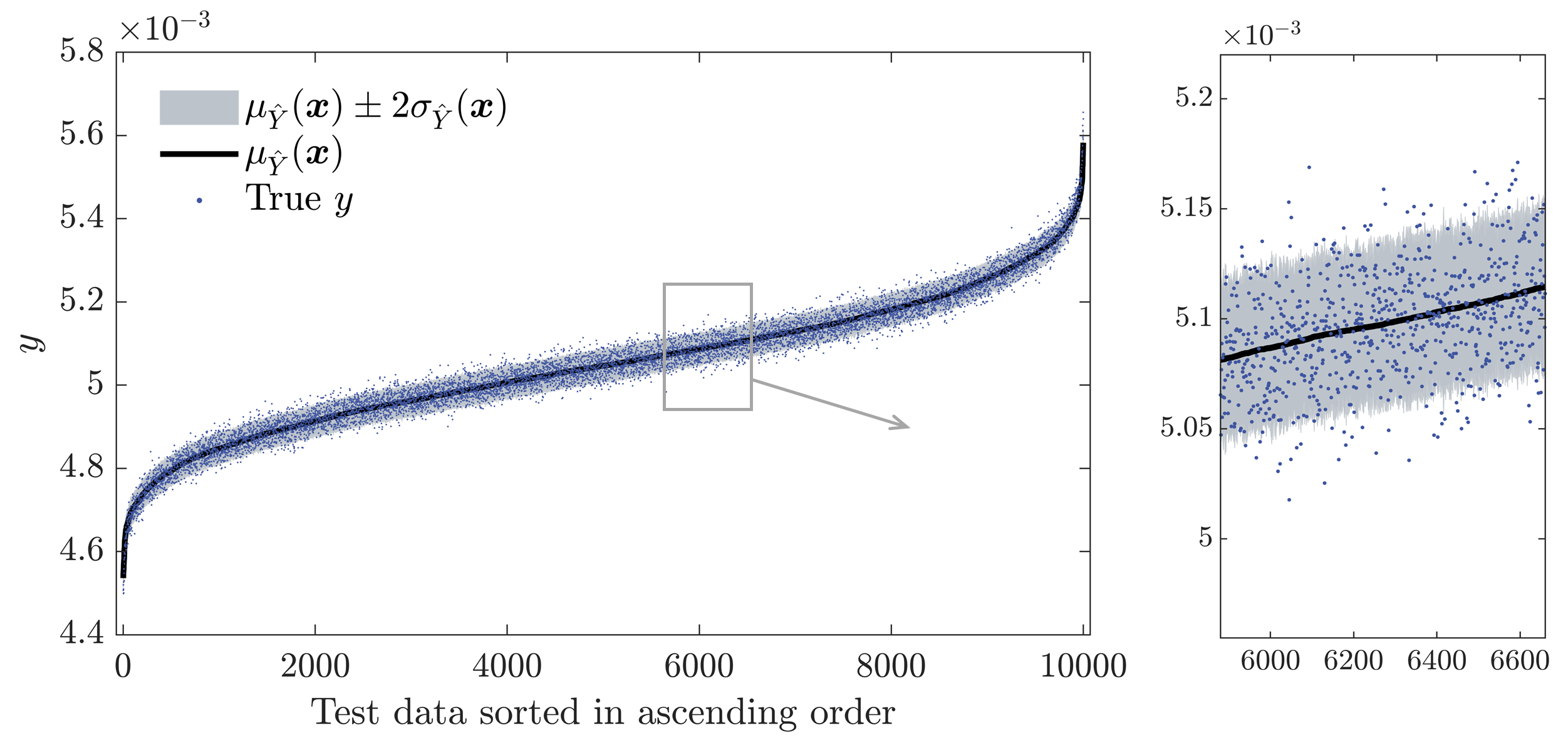}
  \caption{\textbf{A scatter plot of the true model outputs against the surrogate model predictions with uncertainty intervals for the linear elastic bar example.} The true responses are represented by blue circles. Surrogate-based mean predictions and the uncertainty intervals are represented as black line and gray shaded area, respectively.}
  \label{Fig_Ymeanstd_bar}
\end{figure}

Figure \ref{Fig_YYscatter_bar} presents a scatter plot of the true model outputs against the surrogate-based mean predictions, obtained from $10^4$ random test samples. The results show that DR-SM can reproduce the global behavior of the model outputs without significant bias. The figure also presents the mean prediction performance for different reduced dimensions $d$. The results show that the DR-SM algorithm is effective even when the reduced dimensionality is suboptimal, owing to its ability to extract the surrogate model directly from the dimensionality-reduced representation. Although the accuracy gains for increasing $d$ are relatively modest in this particular example, the reduced dimension selection strategy in Algorithm \ref{alg:dimension} serves as a safeguard against underfitting in more complex problems. When the optimal reduced dimension $d=9$ is used, the model achieves the highest predictive accuracy. Figure \ref{Fig_Ymeanstd_bar} presents the predicted mean and uncertainty bounds. The predicted means (black solid line) and standard deviation intervals (gray shaded area) are compared with the true model outputs (blue circles). The result confirms that the mean predictions successfully capture the trend of the true model outputs without overfitting. It is also observed that most of the true responses fall within the two-standard-deviation intervals of the surrogate predictions. Such property is useful in the context of adaptive design of experiments for rare-event probability estimation.

\begin{figure}[H]
  \centering
  \includegraphics[scale=0.425] {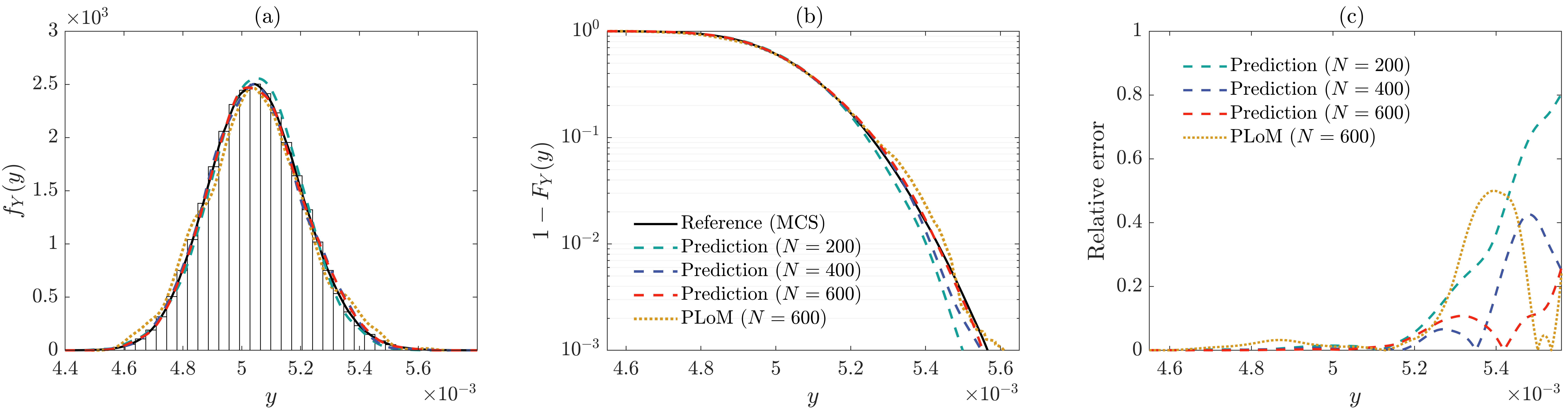}
  \caption{\textbf{Distribution functions estimated by DR-SM, PLoM, and MCS for the linear elastic bar example: (a) PDF, (b) CCDF, and (c) relative errors of the CCDF compared with the MCS reference.} The relative error is defined as the absolute deviation from the MCS reference divided by the MCS value.}
  \label{Fig_pdfcdf_bar}
\end{figure}

Figure \ref{Fig_pdfcdf_bar} illustrates the probability density function (PDF) and complementary cumulative density function (CCDF) obtained from the surrogate model using different number of training samples. The distributions are constructed using mean predictions. A comparison with the MCS and PLoM solutions are also shown in the figure. Details of the PLoM implementation are given in \ref{App:PLoM}. The MCS reference is obtained using $4\times 10^5$ FE simulations, which achieves a coefficient of variation of 5\% at a probability level of $10^{-3}$. To further examine discrepancies at low-probability levels, Figure \ref{Fig_pdfcdf_bar} additionally presents the relative error between the surrogate-based CCDF and the MCS reference. The results indicate that the relative error remains below 15\% for probabilities down to $10^{-3}$ with $N=600$, while it increases for smaller probabilities. Overall, these results confirm that the DR-SM achieves accurate estimation of the response distribution using 600 training samples, resulting in a computational cost reduction to 0.15\% of MCS reference. It is worth reiterating that the proposed surrogate modeling method does not introduce additional tuning parameters other than that in the dimensionality reduction technique and the feature space conditional distribution model.

\begin{figure}[H]
  \centering
  \includegraphics[scale=0.60] {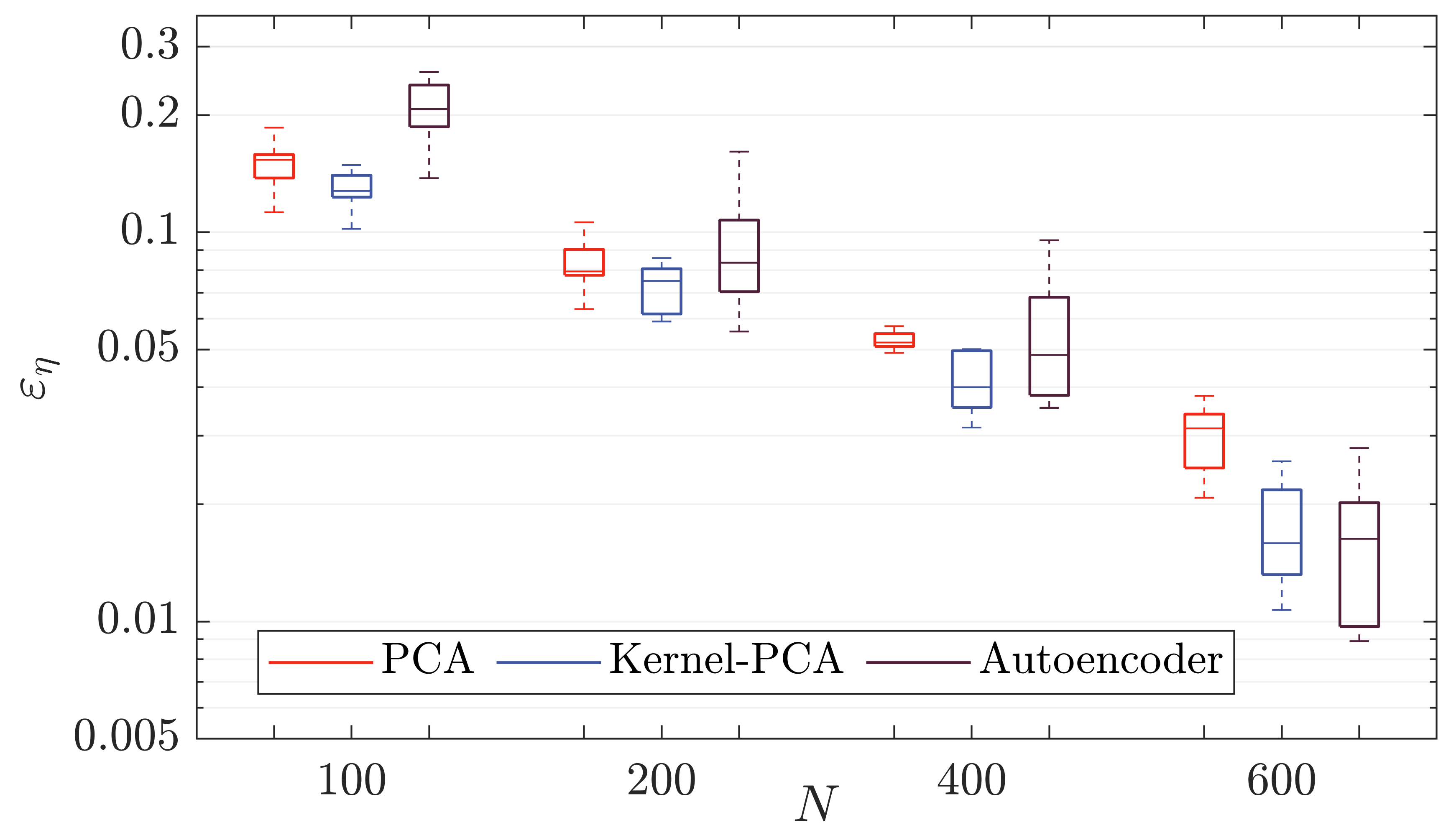}
  \caption{\textbf{Comparison of surrogate model accuracy using different dimensionality reduction techniques with varying training set sizes for the linear elastic bar example.} This plot compares the errors $\varepsilon_{\eta}$ obtained from DR-SM using PCA, kernel-PCA, and autoencoder. Each box plot is obtained using 10 independent runs of the DR-SM algorithm.}
  \label{Fig_error_eta_bar_3methods}
\end{figure}

To compare the accuracy of DR-SM using different dimensionality reduction methods, Figure \ref{Fig_error_eta_bar_3methods} presents the relative mean squared errors $\varepsilon_{\eta}$ associated with PCA, kernel-PCA, and autoencoder. The box plots are obtained using 10 independent runs of DR-SM on different training set sizes $N\in\{100, 200, 400, 600\}$. For the autoencoder model, a fully connected encoder network with 4 hidden layers is used. The encoder compresses the input-output vector into the latent feature space $\vect{\Psi}_{\vect z}$, from which the conditional model is constructed. It is observed that the highest accuracy is achieved with largest training set size, indicating the trade-off between efficiency and predictive performance. All methods are relatively accurate by using $600$ samples.


\subsection{Nonlinear dynamic system under stochastic excitation} \label{BW_example}

\begin{figure}[H]
  \centering
  \includegraphics[scale=0.58] {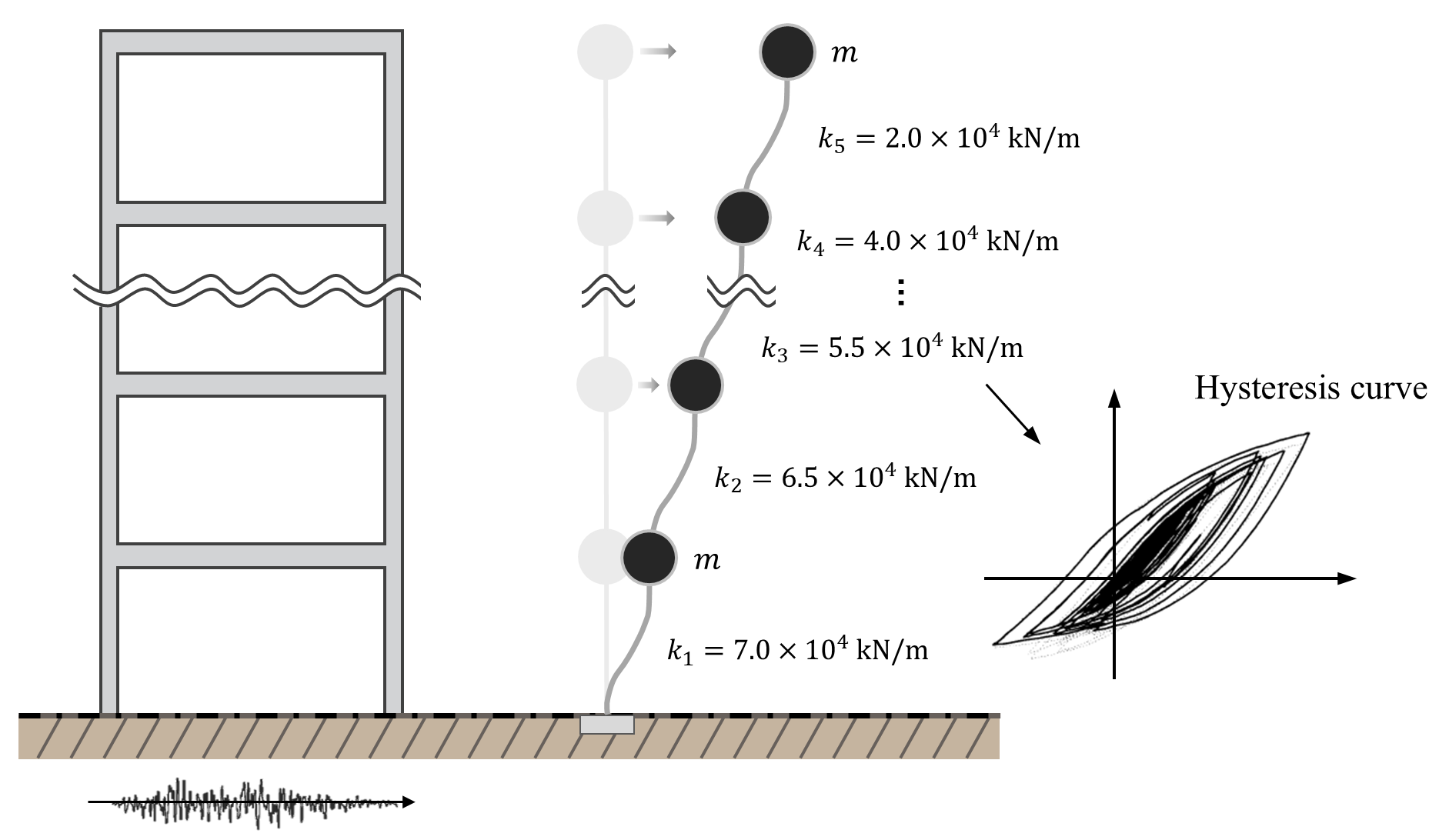}
  \caption{\textbf{Nonlinear dynamical system under stochastic excitation.}} \label{Fig_BW}
\end{figure}

\noindent This example considers a five-story nonlinear hysteretic shear building model under stochastic loading, as illustrated in Figure \ref{Fig_BW}. The system response is obtained by the following differential equation \cite{broccardo2016multicomponent,wang2019hamiltonian,yi2019gaussian}:
\begin{equation} \label{Eq:MDOF_eq}
\vect{M}\ddot{\vect{U}}(t) + \vect{C}\dot{\vect{U}}(t) + \vect{R}(\vect{U}(t),\dot{\vect{U}}(t)) = -\vect{M}\vect{1}\ddot{U}_{g}(t) \,,
\end{equation}
\noindent where $\vect{M}$ and $\vect{C}$ are respectively the mass and damping matrices, $\vect{R}$ is the restoring force function, $\vect{1}$ is a vector of ones, $\vect{U}(t)$, $\dot{\vect{U}}(t)$, and $\ddot{\vect{U}}(t)$ denote the displacement, velocity, and acceleration vectors, respectively, and $\ddot{U}_{g}(t)$ is the ground acceleration process. The in-plane inelastic behavior of the $i$-th story is described through the Bouc-Wen hysteresis model, which is governed by the following set of local differential equations:
\begin{equation} \label{Eq:BW_eq1}
q_i(t) = k_i[\alpha_i v_i(t) + (1-\alpha_i)h_i(t)] \,,
\end{equation}
\begin{equation} \label{Eq:BW_eq2}
\dot{h}_i(t) =  -\delta|\dot{v}_i(t)||h_i(t)|^{(\bar{n}-1)}h_i(t) - \zeta\dot{v}_i(t)|h_i(t)|^{\bar{n}} + A\dot{v}_i(t)\,,  \,\,\,  i=1,...,5 \,, 
\end{equation}
\noindent where $q_i(t)$ is the applied local element force, which in this example is a shear force, $k_i$ is the inter-story elastic stiffness, $h_i(t)$ is the hysteretic response governed by Eq.~\eqref{Eq:BW_eq2}, and $v_i(t)$ denotes the element of the local deformation vector $\vect{v}(t)$ obtained by $\vect{v}(t) = \vect{A}_{\vect{f}}\vect{U}(t)$, where $\vect{A}_{\vect{f}}$ is the compatibility matrix, and $\alpha_i\,, i=1,2,...,5$, are parameters that characterize the degree of inelasticity, with each set to 0.1. The constitutive model parameters are $\bar{n}=3$, $A=1$, and $\delta=\zeta=1/{(2u_{y}^{\bar{n}})}$, in which $u_y=0.01$ m is the yield displacement. A mass of $m=3.0\times10^{4}$ kg is applied to all floors. The damping matrix $\vect C$ is modeled using modal damping, with a 5\% damping coefficient applied to all modes. 

The ground acceleration $\ddot{U}_{g}(t)$ is modeled by a white noise process and discretized in the frequency domain as follows \cite{wang2016cross}:
\begin{equation} \label{WN_eq1}
\ddot{U}_{g}(t) = \sigma\sum_{j=1}^{n/2}\left[X_j\cos{(\omega_{j}t)} + X_{(n/2+j)}\sin{(\omega_{j}t)} \right] \,,
\end{equation}
\noindent where $X_j\,, j=1,...,n$ are independent standard Gaussian random variables. The discretized frequency is given by $\omega_{j}=j\Delta\omega$ with $\Delta\omega=\pi/10$, the cutoff frequency is $\omega_{cut} = 10\pi$ rad/s, $\sigma=\sqrt{2S\Delta\omega}$, and $S=0.015$ m$^2$/s$^3$ is the intensity of the white noise. The discretization is performed with $n=200$, leading to a 200-dimensional input random vector. In this example, the following two response quantities are studied.

\subsubsection{Case 1: Deformation at a given time point}  \label{Example_timpoint}

First, consider the top-story (fifth) deformation at $t=8$ s as the response quantity of interest, denoted as
\begin{equation} \label{Eq:BW_Ymodel1}
Y = \mathcal{M}(\vect X) = U_{top}(8 s; \vect{X}) \,.
\end{equation}

\begin{figure}[H]
  \centering
  \includegraphics[scale=0.55] {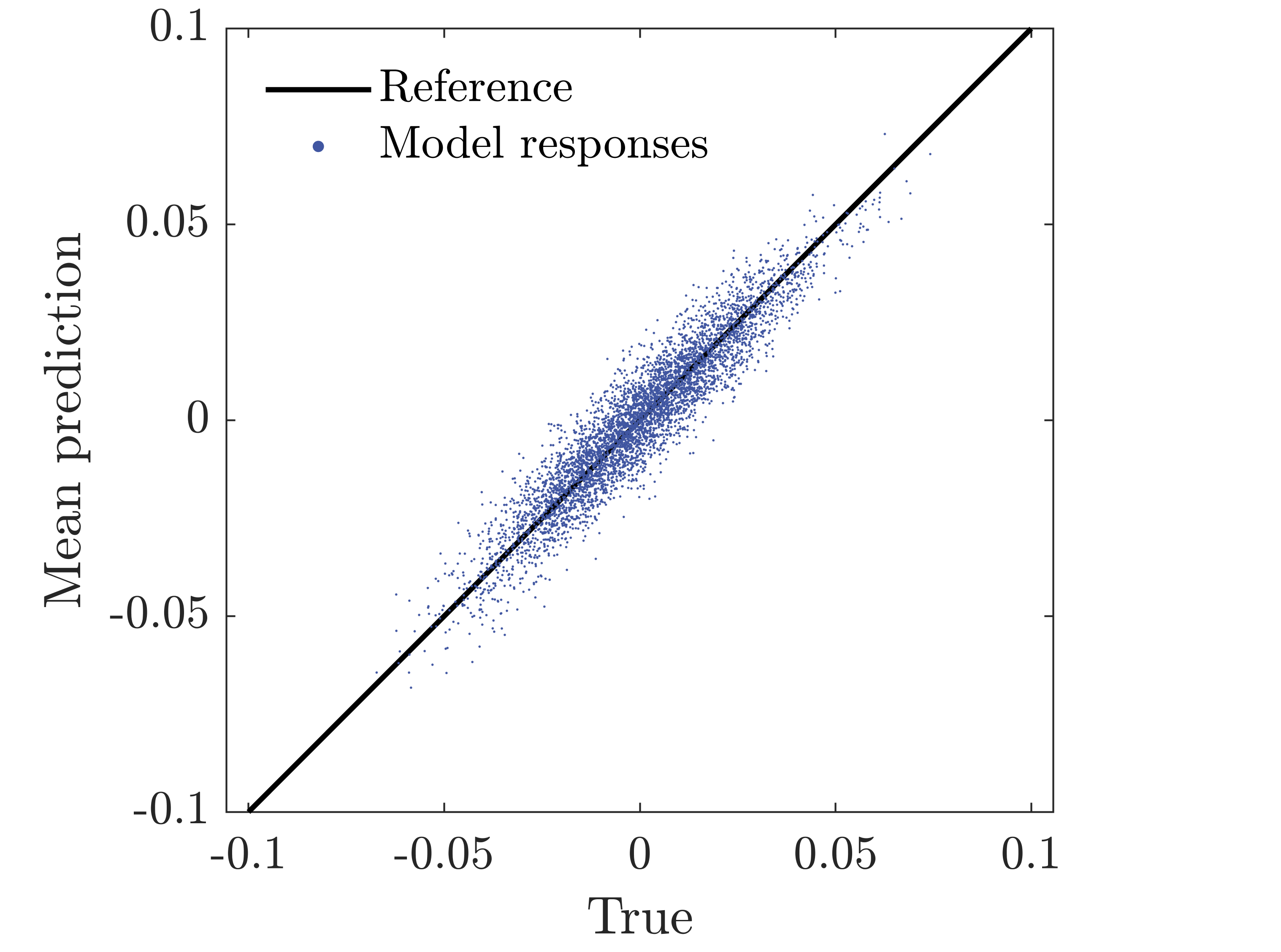}
  \caption{\textbf{A scatter plot of the true model outputs against the surrogate model predictions for the nonlinear dynamic system (Case 1).} The surrogate model is trained using 800 samples, and the relative mean squared error $\varepsilon_{\eta}$ is calculated as $0.1177$.}
  \label{Fig_YYscatter_BW}
\end{figure}

\begin{figure}[H]
  \centering
  \includegraphics[scale=0.425] {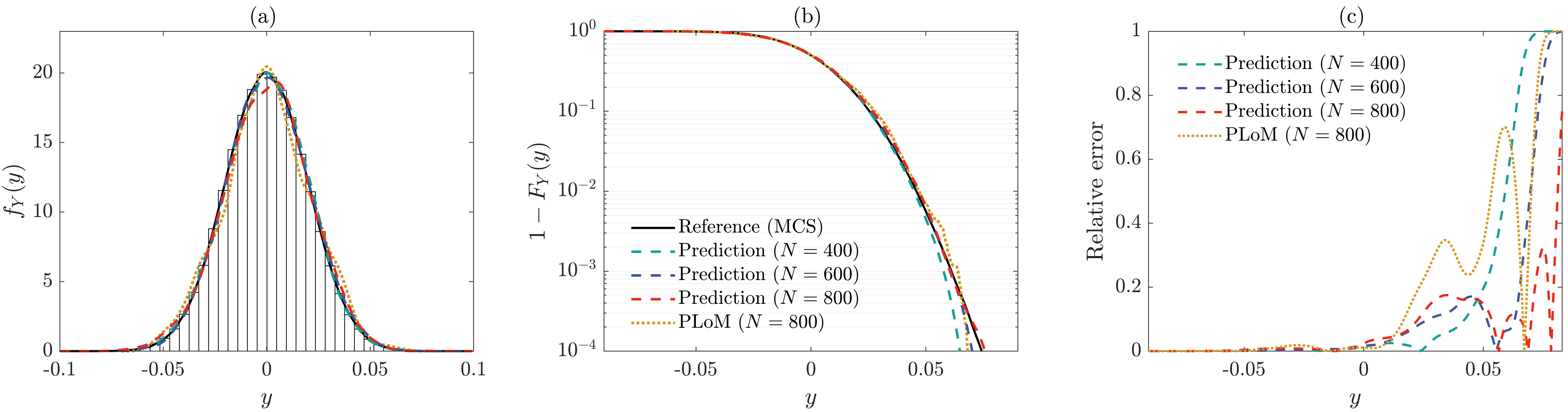}
  \caption{\textbf{Distribution functions estimated by DR-SM, PLoM, and MCS for the nonlinear dynamic system (Case 1): (a) PDF, (b) CCDF, and (c) relative errors of the CCDF compared with the MCS reference.} The relative error is defined as the absolute deviation from the MCS reference divided by the MCS value.}
  \label{Fig_pdfcdf_BW}
\end{figure}

Similarly to Example 2, we first consider the DR-SM using kernel-PCA. The parameters are optimized into $\vect{\theta}^*_{\mathcal{H}} =[1.578,1]$ and $d^*=1$, obtained with a tolerance $\varepsilon_d^t=0.1$. Using 800 training samples, Figure \ref{Fig_YYscatter_BW} presents a scatter plot of the true model outputs against the surrogate-based mean predictions from $10^4$ random test samples. Compared to the previous example, the high nonlinearity present in the hysteric system and the lack of temporal correlation in the input uncertainties, modeled as a Gaussian white noise process, induce larger errors. Figure \ref{Fig_pdfcdf_BW} presents the estimated PDF and CCDF, and their relative errors with respect to the MCS reference, showing that errors remain small for probability levels down to $10^{-3}$, but increase in the lower tail region for smaller probabilities, reflecting the limited coverage of extreme events in the training data. The reference solution is obtained through $4\times 10^5$ dynamic simulations. The results show that the DR-SM using 800 training samples obtains good accuracy in estimating the output distribution function, achieving a cost reduction to 0.2\% of MCS.   


\begin{figure}[H]
  \centering
  \includegraphics[scale=0.60] {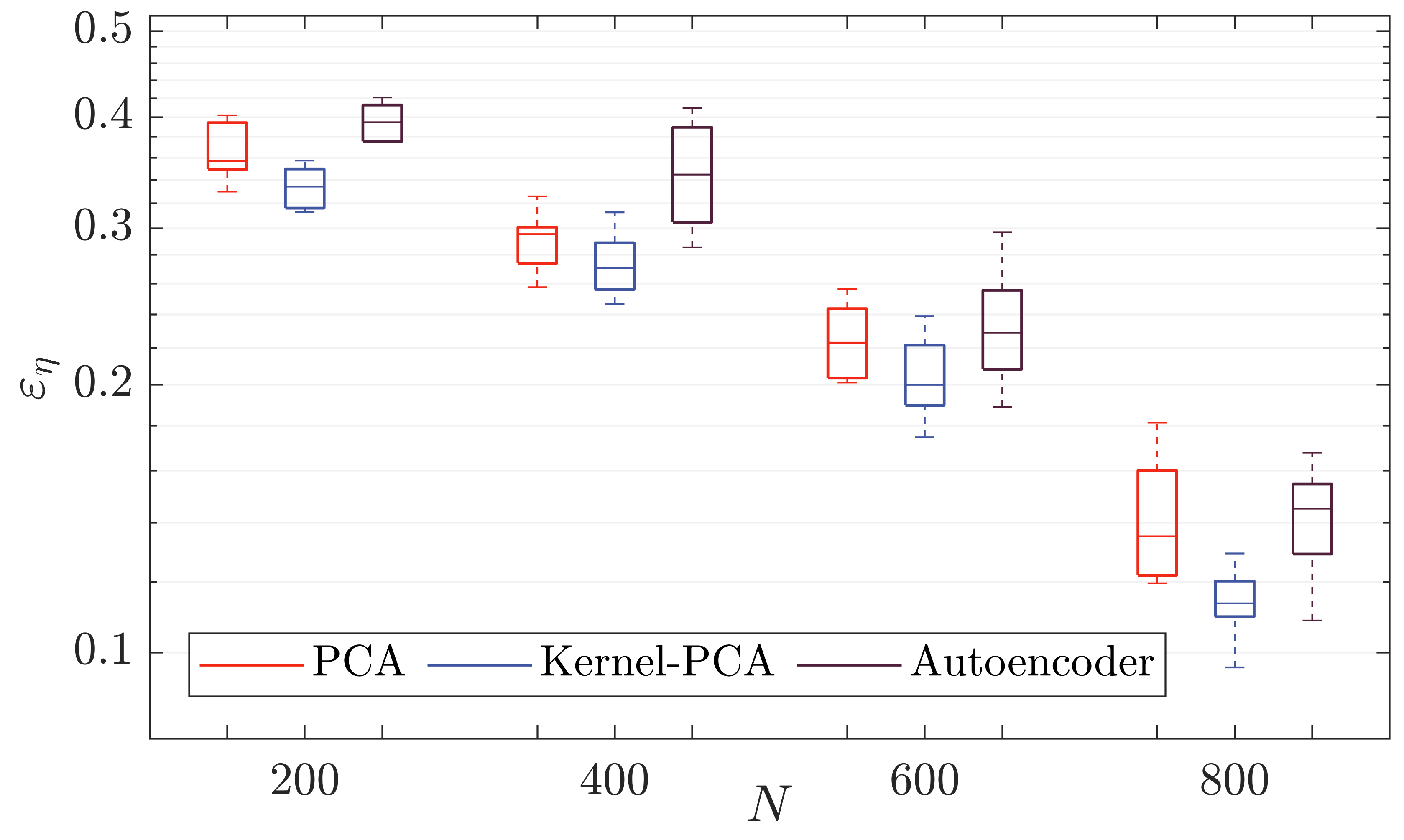}
  \caption{\textbf{Comparison of surrogate model accuracy using different dimensionality reduction techniques with varying training set sizes for the nonlinear dynamic system.} This plot compares the errors $\varepsilon_{\eta}$ obtained from DR-SM using PCA, kernel-PCA, and autoencoder. Each box plot is obtained using 10 independent runs of the DR-SM algorithm.}
  \label{Fig_error_eta_BW_3methods}
\end{figure}

To further examine the effect of training set size and reduced dimensionality, a parametric study is performed. Figure \ref{Fig_error_eta_BW_3methods} compares the relative mean squared errors $\varepsilon_{\eta}$ of DR-SM constructed with PCA, kernel-PCA, and autoencoder for varying training set sizes $N\in\{200, 400, 600, 800\}$. As expected, higher accuracy is achieved with larger training sets. Additionally, Table \ref{Tab_parametric_d} summarizes $\varepsilon_\eta$ for different reduced dimensionalities $d\in\{1, 2, 5, 10, 20\}$, using kernel-PCA and fixed $N=800$. The table reports the mean error values across 10 runs, with standard deviations in parentheses. The results show that prediction errors remain relatively consistent across different reduced dimensions, indicating a lack of clearly dominant low-dimensional structure in the white noise input field, even when augmented with output.

\begin{table}[H]
  \caption{\textbf{Relative mean squared errors of output with varying reduced dimensionalities for the nonlinear dynamic system (Case 1).}}
  \label{Tab_parametric_d}
  \centering
  \begin{tabular}{c c c c c c}
    \toprule
    Reduced dimensionality, $d$ & 1 & 2 & 5 & 10 & 20 \\
    \midrule
    \multirow{2}{*}{Error, $\varepsilon_\eta$} & 0.1159 & 0.1099 & 0.1165 & 0.1270 & 0.1297 \\
     & (0.0106) & (0.0096) & (0.0105) & (0.0132) & (0.0098) \\
    \bottomrule
  \end{tabular}
\end{table}

\subsubsection{Case 2: Peak absolute deformation within a given time interval} \label{Example_FPP}

Next, consider the peak top-story deformation within a given time interval of 10 seconds, defined as:
\begin{equation} \label{Eq:BW_Ymodel2}
Y = \mathcal{M}(\vect X) = \max_{t\in[0,10]} \left|U_{top}(t; \vect X)\right| \,.
\end{equation}
\noindent 
It is often of engineering interest to estimate the distribution of $Y$, or equivalently, the probability of $Y$ exceeding a threshold, referred to as the first-passage probability. In this problem, we observed that standard non-intrusive dimensionality reduction algorithms—such as PCA, kernel-PCA, and autoencoders—were ineffective in capturing a low-dimensional representation, even in the input-output space. Specifically, the iterative simulator failed to converge when these methods were applied, due to instability in the conditional distribution and absence of clear low-rank structure.

To address this, we introduce a physics-based dimensionality reduction using an equivalent linear system, whose peak deformation response serves as an interpretable lower-dimensional representation. In this context, the dimensionality reduction mapping is characterized by parameters of the equivalent linear model. The DR-SM method is then constructed using this physics-guided feature, without requiring additional dimensionality reduction selection procedures. In place of the mapping $\mathcal{H}$, this example introduces a parametric linear system \cite{wang2024optimized,xian2024physics}, defined by the following differential equations:
\begin{equation} \label{Eq:MDOF_linear}
\vect{M}\ddot{\vect{U}}_{eq}(t) + \vect{C}\dot{\vect{U}}_{eq}(t) + \vect{K}_{eq}\vect{U}_{eq}(t) = -\vect{M}\vect{1}\ddot{U}_{g}(t) \,,
\end{equation}
\noindent where $\vect{K}_{eq}$ is the ``equivalent" stiffness matrix composed of inter-story stiffnesses $\vect{k}_{eq}=\{k_{eq_{1}},...,k_{eq_{5}}\}$. The mass and damping matrices are taken to be identical to those of the original nonlinear system in Eq.~\eqref{Eq:MDOF_eq}.

The stiffness parameters $\vect{k}_{eq}^*$ are optimized by maximizing the Pearson correlation $\rho$ between the peak top-story responses of the nonlinear and linear systems:
\begin{equation} \label{Eq:Equilinear_opt}
\vect{k}_{eq}^* = \mathop{\arg\max}_{\vect{k}_{eq} \in \vect{\Omega}_{\vect k}} \rho(Y, Y_{eq}(\vect{k}_{eq})) \,,
\end{equation}
\noindent where $Y_{eq}(\vect{k}_{eq}) = \max_{t\in[0,10]} \left|U_{eq,top}(t; \vect X)\right|$ is peak top-story response obtained from the linear model in Eq.~\eqref{Eq:MDOF_linear}. Through the optimization process, the physics-informed feature space mapping $\Psi_{z}\equiv Y_{eq}(\vect{k}_{eq}^*)$ is obtained. The DR-SM method is then applied to this one-dimensional representation using same conditional distribution modeling and sampling procedures outlined in Section \ref{DRSMdetails}. Using a training set of $N = 1,000$ samples, the stiffnesses are optimized as $\vect{k}_{eq}^*=\{5.62, 6.68, 8.66, 2.98, 1.86\}\times10^4$ kN/m, resulting in a correlation of $\rho=0.8159$.

\begin{figure}[H]
  \centering
  \includegraphics[scale=0.53] {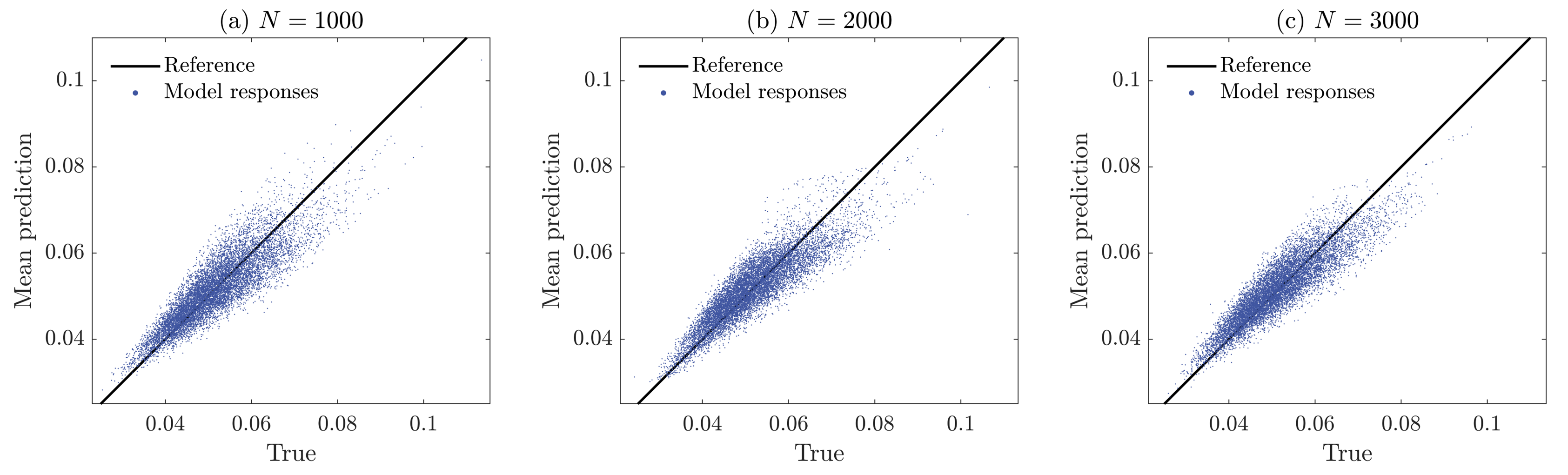}
  \caption{\textbf{A scatter plot of the true model outputs against the surrogate model predictions for the nonlinear dynamic system (Case 2), using training dataset sizes of (a) 1000, (b) 2000, and (c) 3000.} Relative mean squared errors $\varepsilon_{\eta}$ are reported as 0.2342, 0.2158, and 0.1974, respectively.}
  \label{Fig_YYscatter_BWFPP}
\end{figure}

\begin{figure}[H]
  \centering
  \includegraphics[scale=0.425] {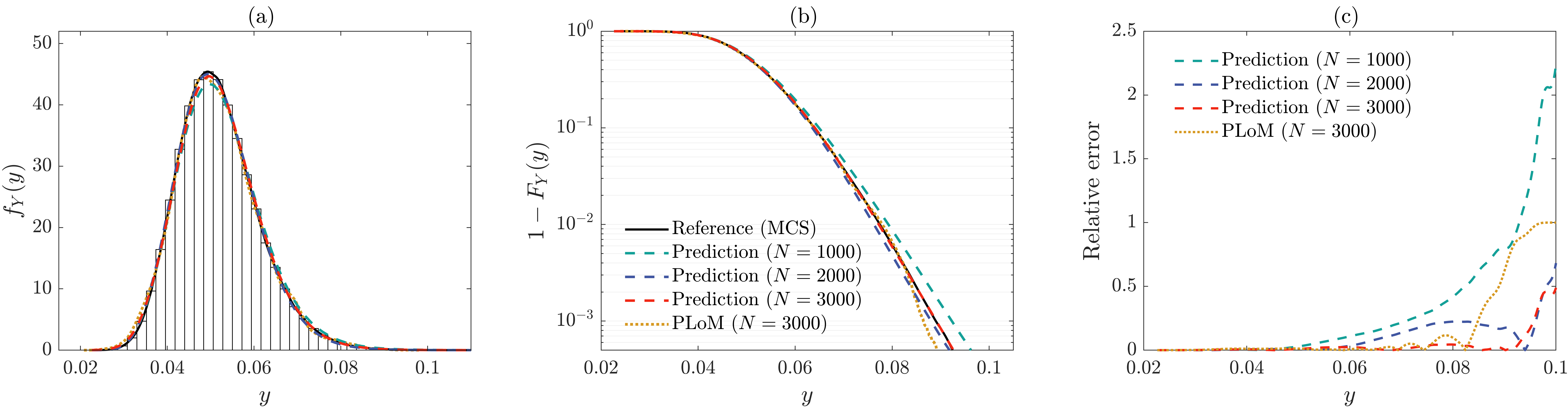}
  \caption{\textbf{Distribution functions estimated by DR-SM, PLoM, and MCS for the nonlinear dynamic system (Case 2): (a) PDF, (b) CCDF, and (c) relative errors of the CCDF compared with the MCS reference.} The relative error is defined as the absolute deviation from the MCS reference divided by the MCS value.}
  \label{Fig_pdfcdf_BWFPP}
\end{figure}

Figure \ref{Fig_YYscatter_BWFPP} presents scatter plots comparing the predicted peak responses with the true values, based on $10^4$ random test samples, for three training dataset sizes: $1000$, $2000$, and $3000$. As the size of the training dataset increases, prediction accuracy improves overall. Nevertheless, considerable variability persists in the high peak response range, reflecting inherent limitations of the linear system regardless of the training dataset size. The accuracy is highest for the largest dataset of $3000$ samples. The estimated distributions of peak absolute deformation are compared with those obtained from PLoM and MCS, as well as their relative errors with respect to the MCS reference, in Figure \ref{Fig_pdfcdf_BWFPP}. The MCS reference is obtained using $4\times 10^5$ dynamic simulations. The results confirm the accuracy of the proposed method.

A final remark on training sample requirements is warranted. In the last example, the proposed method requires around $3000$ training samples to achieve a good accuracy in estimating first-passage probabilities, although accuracy in the high peak response range remains limited due to the use of a linearized system response in the dimensionality reduction step. This might not seem remarkable compared to the efficiency of typical surrogate models in low-dimensional settings. Nevertheless, it is essential to recognize that DR-SM functions as a ``conventional" surrogate model, which predicts the output for a specified input. This feature enables the possibility to integrate with active learning schemes and multi-fidelity UQ methods, offering a pathway to more efficient surrogate modeling in high-dimensional problems. While the current examples focus on numerical studies, we note that the DR-SM model has also been applied to high-fidelity FE models with over 1500 elements and nonlinear hysteretic behavior, demonstrating its practicality for realistic UQ applications involving high-dimensional input \cite{kim2025uncertainty}.


\section{Additional remarks, limitations, and future directions} \label{Remarks}

\subsection{A baseline for testing more complex surrogate modeling methods} \label{Remark:baseline}
The implementation of the proposed surrogate modeling method is straightforward, requiring only (i) a dimensionality reduction algorithm, such as PCA or kernel-PCA, and (ii) a method to model a low-dimensional conditional distribution, such as GP regression or KDE. Beyond the parameters involved in (i) and (ii), the proposed method does not require any additional parameter tuning. Consequently, the proposed method can serve as a baseline for testing more sophisticated surrogate modeling approaches for high-dimensional problems.

\subsection{Heuristic assumption in the proposed method} \label{Remark:heuristic}
The proposed surrogate modeling method hinges on the heuristic assumption that the transition kernel admits a stationary distribution. This assumption is supported by: (i) the correctness in the idealized scenario of exact dimensionality reduction and conditional distribution modeling, and (ii) the success in various numerical examples. In our experience, as long as the dimensionality reduction \textit{works}, i.e., the conditional distribution in the feature space has acceptable accuracy, the stationary distribution exists. Future works are needed to carefully examine the boundary of this heuristic assumption.

\subsection{Physics-informed dimensionality reduction} \label{Remark:physics}
For some highly complex computational models, a physics-informed dimensionality reduction leveraging problem-specific properties may outperform any generic, unsupervised dimensionality reduction techniques. Section \ref{Example_FPP} presents a promising example. Future research could explore other types of problems and work towards identifying universal procedures for developing physics-informed dimensionality reduction techniques.  

\subsection{Rare event probability estimation} \label{Remark:rareevent}
In this paper, the performance of the proposed surrogate modeling method is tested for probability estimations down to $10^{-4}$. While the results demonstrate promising accuracy, the extrapolation capability—particularly in the tail regions of the response distribution—remains limited. To improve prediction accuracy for rare event probability (e.g., $<10^{-5}$), an integration with active learning schemes \cite{kim2021reliability,dang2024semi} can be promising. 

\subsection{Multi-dimensional output problems} \label{Remark:multioutput}

While this paper demonstrates the proposed method for single-output problems, the DR-SM framework can be extended to multivariate output problems by adopting a multivariate conditional distribution model, such as multi-output GP \cite{yi2025multi}, or multivariate KDE. This potential extension warrants further investigation.

\section{Conclusions} \label{Conclusion}
This paper introduces a novel approach for constructing a stochastic surrogate model from the results of dimensionality reduction, named dimensionality reduction-based surrogate modeling (DR-SM). DR-SM aims to address uncertainty quantification problems characterized by high-dimensional input uncertainties and computationally intensive physics-based models. This method is versatile, not confined to any particular dimensionality reduction algorithm, and is straightforward to apply. The core of DR-SM is a transition kernel that alternates between a dimensionality reduction algorithm and a feature space conditional distribution. The performance of DR-SM is validated through three high-dimensional uncertainty quantification problems, including a first-passage probability estimation problem known to be challenging for dimensionality reduction-based methods. The proposed method exhibits good accuracy for the tested problems. Future research efforts can target (i) rare-event simulation, (ii) multivariate output problems, and (iii) physics-informed dimensionality reduction. 

\section*{Acknowledgments}
This research was supported by the Pacific Earthquake Engineering Research Center grant NCTRZW. The first author was supported by the Basic Science Research Program through the National Research Foundation of Korea (NRF) funded by the Ministry of Education (RS-2024-00407901). 

\section*{Data availability}
The source codes are available for download at \url{https://github.com/Jungh0Kim/DR-SM}.

\appendix
\renewcommand{\theequation}{A.\arabic{equation}}
\renewcommand{\thefigure}{A.\arabic{figure}}
\renewcommand{\thetable}{A.\arabic{table}}
\setcounter{figure}{0} 
\setcounter{table}{0} 

\section{PLoM implementation}\label{App:PLoM}

This appendix outlines the implementation steps of the Probabilistic Learning on Manifolds (PLoM) method \cite{soize2016data,soize2019entropy,soize2020physics} as used in this study.
\begin{itemize}
    \item \textbf{Data normalization}: All input-output samples are linearly scaled so that each variable lies within the interval $[0,1]$, preserving invertibility for post-processing. The scaling coefficients and mean vector are stored to recover the original physical space.
    \item \textbf{Decorrelation via PCA}: PCA is applied to the normalized data matrix to decorrelate the data points.
    \item \textbf{Non-parametric density estimation}: A KDE of the decorrelated data is constructed using a Gaussian kernel with bandwidth selected automatically (via Silverman’s rule). This provides an approximation of the latent space distribution.
    \item \textbf{Diffusion maps and basis construction}: A diffusion kernel is applied to the reduced data to capture the underlying manifold geometry. A reduced set of basis vectors is obtained from the leading eigenvectors of the diffusion operator, enabling nonlinear structure-preserving projection.
    \item \textbf{It\^{o} stochastic differential equation (ISDE) sampling}: A dissipative Hamiltonian system is formulated to sample from the latent KDE. The ISDE is solved using a second-order St\"{o}rmer--Verlet scheme, with trajectories projected onto the diffusion basis.
    \item \textbf{Back-projection and unscaling}: The generated samples in the latent space are mapped back to the normalized space via inverse PCA, and then unscaled to the original physical space using the stored normalization parameters.
\end{itemize}

\renewcommand{\theequation}{B.\arabic{equation}}

\section{Bayesian inference in hGP model} \label{App:hGP}

\noindent This appendix summarizes the Bayesian inference procedure for conditional distribution modeling using the heteroscedastic Gaussian process (hGP) model \cite{lazaro2011variational} employed in this study. The hGP model introduces an input-dependent Gaussian noise term:
\begin{equation}  \label{Eq:hGP_model}
y(\vect{x}) = g(\vect{x}) + \varepsilon(\vect{x})\,, \,\,\,\,\,\,\,\, \varepsilon(\vect{x})\sim
N(0, \exp{(r(\vect{x}))})\, ,
\end{equation}
where both the latent output function $g(\vect{x})$ and the noise variance function $r(\vect{x})$ follow Gaussian process prior with respective kernel hyperparameters $\vect{\theta}_g$ and $\vect{\theta}_r$.

To estimate the hyperparameters for latent functions $g(\vect{x})$ and $r(\vect{x})$, a variational Bayesian approach is adopted. The inference is based on maximizing a tractable lower bound of the marginal likelihood:
\begin{equation} \label{Eq:hGP_MLE}
b_{MV}(\vect{m},\vect{V}) = \ln f_N(\mathcal{Y}_\mathcal{D};\vect{0},\vect{K_g}+\vect{Q}) - \frac{1}{4}tr(\vect{V}) - KL\left(f_N(\vect{r};\vect{m},\vect{V}) || f_N(\vect{r};\mu_{0}\vect{1},\vect{K_r})\right)    \,,
\end{equation}
where $\mathcal{Y}_\mathcal{D}$ is observations, $KL(\cdot)$ denotes the Kullback-Leibler divergence, $\mu_{0}$ is mean kernel for $r(\vect{x})$, $\vect{K_g}$ and $\vect{K_r}$ are the kernel matrices, and $\vect{Q}$ is a diagonal matrix with elements $Q_{i,i} = \exp(m_i - V_{i,i}/2), \, i=1,...,N$.

Given the optimized hyperparameters and variational parameters $\vect{m}$ and $\vect{V}$, the predictive distribution at a test point $\vect{x}_*$ is obtained by marginalizing over the latent functions. The predictive mean and variance of the distribution for $y$ can be derived as:
\begin{equation}  \label{Eq:hGP_mean}
\mu_{Y}({\vect{x}_*}) = \vect{k}_{\vect{g}_*}^T(\vect{K_g}+\vect{Q})^{-1}\mathcal{Y}_\mathcal{D} \,,
\end{equation}
\begin{equation}  \label{Eq:hGP_sigma}
\sigma_{Y}^2({\vect{x}_*}) = \exp(\chi_* + \eta_*^2/2) + k_g(\vect{x}_{*},\vect{x}_{*}) - \vect{k}_{\vect{g}_*}^T (\vect{K_g} + \vect{Q})^{-1} \vect{k}_{\vect{g}_*}  \,.
\end{equation}
where $\chi_*=\vect{k}_{\vect{r}_*}^T(\vect{\Lambda}-\frac{1}{2}\vect{I})\vect{1}+\mu_0$ and $\eta_*^2=k_r(\vect{x}_{*},\vect{x}_{*})-\vect{k}_{\vect{r}_*}^T(\vect{K_r}+\vect{\Lambda}^{-1})^{-1}\vect{k}_{\vect{r}_*}$ are the predictive mean and variance for $r(\vect{x})$, $\vect{\Lambda}$ is a diagonal matrix used to reparameterize the variational parameters $\vect{m}$ and $\vect{V}$ in a reduced order, $\vect{k}_{\vect{g}_*}$ and $\vect{k}_{\vect{r}_*}$ are covariances between prediction location and training points for $g(\vect{x})$ and $r(\vect{x})$, respectively.


\bibliography{DRSM}


\end{document}